# Circuit-level-configurable Zero-field Superconducting Diodes: A Universal Platform Beyond Intrinsic Symmetry Breaking

Xiaofan Shi[1, 4, †], Ziwei Dou[1, †,*], Dong Pan[2,†], Guoan Li[1, 4, †], Yupeng Li[1,5], Anqi Wang[1],

Zhiyuan Zhang[1, 4], Xingchen Guo[1, 4], Xiao Deng[1, 4], Bingbing Tong[1], Zhaozheng Lyu[1], Peiling

Li[1], Fanming Qu[1, 4, 6, 7],Guangtong Liu[1, 6, 7], Jianhua Zhao[2, 3*], Jiangping Hu[1, 8, 9*], Li Lu[1, 4, 6, 7*],

Jie Shen[1, 6*]

[1]Beijing National Laboratory for Condensed Matter Physics and Institute of Physics, Chinese

Academy of Sciences, Beijing 100190, China

[2]State Key Laboratory of Semiconductor Physics and Chip Technologies, Institute of

Semiconductors, Chinese Academy of Sciences, Beijing 100083, China

[3]National Key Laboratory of Spintronics, Hangzhou International Innovation Institute,

Beihang University, Hangzhou 311115, China

[4]University of Chinese Academy of Sciences, Beijing 100049, China

[5]Hangzhou Key Laboratory of Quantum Matter, School of Physics, Hangzhou Normal

University, Hangzhou 311121, China

[6]Songshan Lake Materials Laboratory, Dongguan, Guangdong 523808, China

[7]Hefei National Laboratory, Hefei 230088, China

[8]Kavli Institute of Theoretical Sciences, University of Chinese Academy of Sciences, Beijing,

100190, China

[9]New Cornerstone Science Laboratory, Shenzhen 518054, China

* Corresponding authors. E-mails: ziweidou@iphy.ac.cn, jhzhao@semi.ac.cn,

jphu@iphy.ac.cn, lilu@iphy.ac.cn, shenjie@iphy.ac.cn,

†These authors contributed equally to this work.

**Abstract**

Modern industry seeks next-generation microelectronics with ultra-low dissipation and noise beyond semiconducting systems, where the superconducting electronics offer promise. Its physical foundation is the superconducting diode effect (SDE) with nonreciprocal supercurrent. SDE has hitherto mainly relied on material-specific intrinsic symmetry breaking in superconductors, suffering from low yield, controllability, and compatibility with further functional extension - an undesirable aspect for applications. Here, we demonstrated a field-free SDE due to the chemical potential shift from external circuit line resistance, which is generic and challenges the previous interpretations of the intrinsic symmetry breaking in superconductivity for zero-field SDE. Moreover, this SDE is circuit-level configurable since it can be electrically switched on/off with its polarity and efficiency precisely modulated via gate voltage and circuit reconfiguration, facilitating functional extension. Such a generic, controllable and extensible SDE addresses critical challenges in dissipationless circuit towards application, and thus establishes a robust platform for scalable superconducting electronics.

## Main text

Microelectronic circuitry is one of the cornerstones for modern industry. The gigantic computational power and quantum-limited precision necessary for many emerging technologies pose significant challenge for such circuitry with ultra-low dissipation and noise beyond the prevailing semiconductor-based systems. Superconducting electronics thus comes to the forefront due to its intrinsic dissipationless supercurrent and its potentials in future applications of aerospace technology and quantum information and sensing, etc.[1-3]. Similar to the semiconducting circuitry, the building block of such superconducting electronic system is the superconducting diode (SDE) with unequal supercurrent between positive and negative directions[4-9]. While the SDE holds great promise for applications in quantum technologies, its practical implementation requires precise control, scalability, and seamless integration in order to construct low-temperature electronic systems such as superconducting logic chips.

To realize the superconducting diode effect, several symmetries should be simultaneously broken[4-9], namely, the inversion symmetry breaking (ISB) and the time-reversal symmetry breaking (TRSB). ISB is required as in the semiconducting diodes[10]. TRSB, however, is further required in SDE, since zero resistance from superconductivity imposes zero-voltage drop and thus ISB alone is not enough to create nonreciprocal supercurrent[4-9]. Since its prediction[4], many systems have achieved such SDE, both in intrinsic superconductors[11-13] and in hybrid junctions[14-23]. The ISB is realized by strong spin-orbit interaction[14-16], asymmetric interface[17] or device structure[21], or noncentrosymmetric lattice[11,19] etc. The TRSB, meanwhile, is most commonly induced by external magnetic field[11,14-17,21,23], which is undesirable for practical applications. "Zero-field" SDE, which does not require external field, thus has attracted intensive research interest recently. Such zero-field SDE is typically achieved by ferromagnetism[18-20] or other exotic spontaneously polarized spin/valley mechanisms[12,13], while in a few other systems the TRSB mechanisms are still controversial[24-29]. To date, what the fundamental mechanisms underlying the "zero-field" SDE are is still unresolved and is one of the pivotal challenges in this field. Moreover, since these SDE systems are realized by material-specific and complicated intrinsic ISB/TRSB mechanisms of the superconducting systems themselves, the inhomogeneous material quality and the fragile electronic state of the superconductors usually results in insufficient reproducibility and controllability of SDE. This may limit the advanced applications, which require robust SDE with controllable polarity and efficiency as well as the flexibility for engineering the extended functionalities. Therefore, searching for a SDE with generic, controllable, and extensible features are highly in demand.

Here, we report a paradigm-shifting field-free SDE not caused by material-specific intrinsic ISB/TRSB in the superconducting device, but by the chemical potential shift from the external circuit line resistance $R_L$ which is equally indispensable for constructing the superconducting diode. This platform-independent approach achieves: generality since it is independent of material-specific ISB/TRSB mechanisms of the superconducting device; controllability since its polarity and efficiency can be readily configured and optimized by $R_L$ assisted with gate-tuning; and extensibility since further functional engineering can in principle be done in the external circuit without the stringent limitation from the low-temperature environment as well as the fragile superconducting materials. By harnessing a superconductor-semiconducting nanowire hybrid Cooper-pair transistor (CPT) as a prototype, we demonstrated such generic, controllable and extensible SDE operating at zero field. First, by applying magnetic field, $\Delta I_C$ did not change sign and was an even-function of $B$, confirming its zero-field nature. Second, we also found the SDE polarity can be switched by the sign of the chemical potential shift, while the SDE can be eliminated altogether if the chemical potential shift was actively tuned to zero. Finally, we provided a controllable and extensible way to tune the diode efficiency in a wide range and achieve the maximal efficiency of ~ 60 % by configuring $R_L$ and $I_C$, which can be done at room temperature assisted by gate-tuning. Our work therefore highlights the hitherto less-recognized importance of the external circuitry besides the intrinsic symmetry breaking mechanism in designing SDE. It not only suggests that interpretations for intrinsic symmetry breaking, in particular TRSB, based on exotic zero-field SDE in superconductivity need to be re-examined in light of this chemical potential shift mechanism, but also opens up a circuit-level configurable and thus more reliable route of developing superconducting electronics into more complex architecture necessary for advanced applications.

**Generic zero-field SDE by chemical potential shift**

The superconducting device can be modeled as a two-terminal device with the source (S) and drain (D) contacts. Ideally S is connected to a dc current source while D is connected to ground directly. However, in realistic setup, the measurement lines connecting the room-temperature circuitry and the device at low-temperature generally have line resistance $R_L$ varying from ten- ohms to kilo-Ohms, which are indispensable for low-temperature and low-noise measurement due to various filters ($R_L$)[30]. In particular, if we include $R_L$ between D and the ground (Fig. 1a), the actual chemical potential of the device is then shifted[31,32] by -|e|$I_{DC}R_L$ when $I_{DC}$ is applied (e is the charge of electron). Taking the system with its original Fermi level lying in the conduction band as an example in Fig. 1b. For positive $I_{C+}$, a negative chemical potential shift  -|e|$I_{C+}R_L$ decreases its Fermi level to $E_{F+}$ and also the electron carrier density in the system. This in turn changes the critical current to $I_C(E_{F+})$. Similarly, for negative $I_{C-}$, the positive -|e|$I_{C-}R_L$ increases its Fermi level to $E_{F-}$ and thus changes the critical current to $I_C(E_{F-})$. If $I_C$ is sensitive to $E_F$ of the order of |e|$I_CR_L$ which is possible for single-electron/Cooper-pair devices[33], low dimensional materials[31,32,34], or superconductors with high $T_C$[35], such chemical potential shift caused by the external $R_L$ in the circuit results in a different

$I_{C+} = I_C(E_{F+})$ and $|I_{C-}| = |I_C(E_{F-})|$, and thus a SDE which does not depend on intrinsic ISB/TRSB mechanism of the superconducting device (Fig. 1c).

In order to demonstrate such SDE induced by chemical potential shift, we choose the single Cooper-pair device called Cooper-pair transistor (CPT, Fig. 1d) whose behavior is sensitive to chemical potential shift. Also its high gate controllability (as will be explained below) may be designed by the geometry-related charging energy instead of depending on material-specific details, which facilitates generic design of the superconducting diodes and further functional electronic systems. The CPT is a well-understood quantum device widely used in constructing transmon qubit[36-38] and detecting Majorana zero modes in hybrid systems [39,40]. It is formed by a small hybrid superconducting island with the charging energy $E_C$, which is further coupled to the superconducting S and D via tunnel junctions with the Josephson energy $E_J$. The charge number of the CPT and its chemical potential can be further fine-tuned by the local gate $V_g$ capacitively coupled to the island. The hybrid configuration allows the two local gates to tune the tunneling barriers, and thus the $E_C$ and $E_J$. For $E_C \sim E_J$, the Coulomb blockade effect is significant and its $I_C$ oscillates with $V_g$ [36-38]. A typical $I_C(V_g)$ is illustrated in Fig. 1f (black dashed curve in the top panel). $I_C$ is maximal if $V_g$ is such that the Fermi level of the island is aligned with those of the S and D ("on Coulomb resonance"), while it is minimal if the Fermi level of the island is misaligned ("Coulomb blockaded"). The average $I_C$ is determined by $E_J$ while the amplitude of the oscillation increases with $E_C/E_J$ [36-38] (see Supplementary Information Sec. 6 for detailed behaviors of CPT). Due to such Coulomb blockade effect, $I_C$ thus has significant response to energy shift around $E_C$ making it sensitive to the chemical potential shift by $-|e|I_{DC}R_L$. Therefore, if $V_g$ is such that the island is charge neutral, for positive $I_{C+}$, the chemical potential decreases, causing the island to be slightly hole-doped (Fig. 1e). A positive $V_g$ inducing electrons is thus required to restore the charge neutrality, resulting in the entire $I_{C+}(V_g)$ shifted in the positive direction of the $V_g$ axis (red solid curve in Fig. 1f, top panel). Conversely, the negative $I_{C-}$ shifts $|I_{C-}(V_g)|$ in the negative direction of the $V_g$ axis (blue solid curve in Fig. 1f, top panel). Such bias-dependent $-|e|I_{C+,-}R_L$ thus results in a gate-tunable SDE at zero-field (Fig. 1f, bottom panel) without TRSB/ISB intrinsic to the CPT.

**Observation of zero-field and gate-switchable SDE in CPT**

Fig. 2a shows the SEM image of the CPT based on InAs nanowires[41,42] (Device 1, simplified as D1). The contacts (dark blue) are formed by depositing the Ti/Al bilayer of 5 nm/65 nm. The central section of the nanowire is covered with ~ 8 nm epitaxially grown Al (pale blue) to form the Cooper pair island, whose Fermi level and $E_{C, J}$ are controlled by the plunger gate ($V_{PG}$) and the two tunnel gates ($V_{TG1,2}$), respectively. A global backgate ($V_{BG}$) via the doped silicon substrate with $SiO_2$ layer can also be applied to control the doping of the entire nanowire (Fig. 2b). Details for material preparation and fabrication are included in Supplementary Information Sec. 1. The superconducting device is biased with $R_L$ = 11 k$\Omega$ connected to D and is measured using standard lock-in amplifier technique (details see Supplementary Fig. 1a). All the measurements are performed at zero magnetic field and the base electronic temperature below 10 mK unless stated otherwise. Fig. 2c (top panel) shows the differential resistance $R$ = d$V$/d$I$ versus $V_{PG}$ and $I_{DC}$. For each $V_{PG}$, $I_{DC}$ is always swept from 0 (superconducting state) to finite positive/negative values (finite-voltage state) (see Supplementary Fig. 2 for sweeps from finite-voltage state to superconducting state[43]). $I_{C+,-}(V_{PG})$ is thus the boundary between the zero-resistance and finite-resistance states. By tuning $V_{TG1,2}$ such that $E_C \sim E_J$, $I_{C+,-}(V_{PG})$ both oscillate periodically as a result of the Coulomb blockade effect[36], from which we estimate $E_C \approx 41$ $\mu$eV and $E_J \approx 80$ $\mu$eV (see Supplementary Figs. 9, 10 for details) comparable to previous similar works[37,38,40]. However, remarkably, $I_{C+}(V_{PG})$ and $|I_{C-}(V_{PG})|$ (the red and blue curves in the middle panel of Fig. 2c, respectively) do not coincide, with the positive (negative) branch shifted in the positive (negative) direction of the $V_{PG}$ axis. As a result, a non-zero $\Delta I_C = I_{C+} - |I_{C-}|$ (black curve, bottom panel of Fig. 2c) appears and also strongly oscillates with $V_{PG}$, and we thus observed a gate-dependent SDE as described in Fig. 1f. Interestingly, $\Delta I_C$ can be tuned either finite or zero (green dashed line in Fig. 2c and Fig. 2d with green square) and its sign can be reversed by gate (red/blue dashed lines in Fig. 2c and Fig. 2d with red/blue square), meaning the presence of the SDE as well as its efficiency and polarity can be switched simply by gate. By taking the linecuts $R(I_{DC})$ along the dashed lines in Fig. 2c (shown in Fig. 2d), we confirm zero-voltage drop before $I_C$ and thus excludes TRSB due to residual electrical field/voltage across the CPT[4-9,43].

Furthermore, $\Delta I_C$ appears in the absence of magnetic field. To confirm its zero-field nature, we fix $V_{PG}$ at the most negative $\Delta I_C$ and measure the dependence of $R$ versus $I_{DC}$ and $B$ (Fig. 2e). The extracted $I_{C+}$ and $|I_{C-}|$ shows maximal $-\Delta I_C$ at $B = 0$ which is suppressed as an even function[8,24,25,28,29] of $B$ (Fig. 2e, bottom panel). Here $B$ is paralleled to $I_{DC}$, and similar behavior is also observed when $B$ is in other directions (see Supplementary Figs. 5a-c) and also in another similar Device 2 (D2, see Supplementary Fig. 4 and Supplementary Figs. 5d-f). This is qualitatively different from the finite-field SDE where $\Delta I_C = 0$ at $B = 0$, as well as is an odd function with $B$ and changes sign at finite positive/negative field[11,14-17,21,23]. Such $\Delta I_C(B)$ is also different from the zero-field SDE with ferromagnetism, whose $\Delta I_C(B)$ is maximized at finite $B$ due to coercive field and also changes sign with $B$ [18-20]. Besides, $\Delta I_C$ shows no visible hysteresis with $V_{PG}$ (see Supplementary Figs. 3f, g), which also excludes spontaneously polarized spin/valley at certain Fermi levels[12,13]. Similar zero-field SDE without external magnetic field, ferromagnetism, or other spontaneous TRSB mechanism has been reported in previous works which attempt to explain the zero-field SDE in terms of more exotic mechanisms[24-29] etc. However, those mechanisms, which are highly specific to the materials, are unlikely to be applicable to the CPT based on InAs semiconducting nanowires.

**Chemical potential shift as the cause for zero-field SDE**

Indeed, in the following, we are able to explicitly exclude the intrinsic TRSB/ISB of CPT, as well as to confirm that the material-independent chemical potential shift as the cause for the observed zero-field SDE. Since the CPT alone in Fig. 2 (D1) requires small $I_C$ in order to have significant $E_C/E_J$ [36-38] for oscillating $I_C$, we fabricated another Device 3 (D3) consisting of similar CPT in parallel with a simple Al-InAs nanowire Josephson junction (JJ) to form a superconducting loop. This enables additional increase of $I_C$ via the paralleled JJ by tuning the side-gate voltage $V_{SG}$ without significantly affecting $E_C/E_J$ of the CPT, thus further enhancing the chemical potential shift $-|e|I_C R_L$. Figs. 3a, b show the device image and its schematic measurement setup. In order to boost measurement efficiency, we use the fast counter measurement technique which applies repeated dc current pulses and detects $I_{C+}$ (or $|I_{C-}|$) in short time. Such technique obtains the same $I_{C+,-}$ as the conventional lock-in technique but with much greater speed, and is widely accepted in $I_C$ and SDE measurements [15,37,38,44] (see Supplementary Figs. 1b, c for more details).

When D is grounded via $R_L$ (Fig. 3c), Fig. 3d reproduces the similar zero-field SDE observed in Fig. 2c with $I_{C+}(V_{PG})$ shifted relatively to higher $V_{PG}$ than $|I_{C-}(V_{PG})|$ and the gate-switchable $\Delta I_C$. The plot is measured with $I_{DC}$ first sweeping in positive direction and then the negative direction. By swapping the sweeping order (first 0 to negative then 0 to positive values), both $I_{C+,-}(V_{PG})$ and $\Delta I_C(V_{PG})$ remain unchanged (Fig. 3e). Therefore if any TRSB is present in the device, its polarity is not initialized by the direction of the initial current bias [27]. It also excludes the possibility of time-dependent drift as the cause of the zero-field SDE. In Fig. 3f, we further swap the contact to be grounded via $R_L$ from D to S. This time, however, the measured $I_{C+}(V_{PG})$ in Fig. 3g is now shifted relatively to lower $V_{PG}$ than $|I_{C-}(V_{PG})|$, and $\Delta I_C$ becomes opposite to Fig. 3d. It is clearly that the $I_{DC}(V_{PG})$ with $R_L$ connected to the ground is always shifted to the higher $V_{PG}$ while that with $R_L$ connected to the current source is always shifted to the lower $V_{PG}$. The polarity switching of the SDE simply due to choosing whether S or D contact to be grounded strongly indicates that SDE is not due to intrinsic TRSB/ISB of the CPT system, since the SDE polarity should remain the same under the swapping operation if there is fixed intrinsic TRSB/ISB which is not switchable by initial current bias as proved in Fig. 3e. Meanwhile, such polarity switching by swapping the grounded contact agrees with the chemical potential shift scenario, since the chemical potential shift in Fig. 3f configuration is $+|e|I_{DC}R_L$ and has the opposite sign of $-|e|I_{DC}R_L$ in Fig. 3c. We emphasize that other line resistance between the current source and the CPT (drawn in Figs. 3(C, F)) does not affect its chemical potential.

We are able to further confirm the chemical potential shift scenario by intentionally eliminating such effect with the "symmetric biasing" scheme illustrated in Fig. 3h. Here S is connected to $I_{DC}$ flowing into S while D is connected to $I_{DC}$ flowing away from D. The currents provided by the two sources are carefully correlated, checked by the high-precision digital oscilloscope. As a result, the chemical potential shift on S and D cancels each other, leaving zero chemical potential shift of the device. Indeed, with such symmetric biasing scheme, the relative shift between $I_{C+}(V_{PG})$ and $|I_{C-}(V_{PG})|$ disappears and no SDE is observed (Fig. 3i). The above measurements together prove unequivocally the role of the generic chemical potential shift as the cause of zero-field SDE instead of any intrinsic and material-specific ISB/TRSB mechanism of the superconducting device. Similar tests are also reproduced in D1 with lock-in amplifier technique (Supplementary Figs. 3a-e).

**Control and extension of the generic zero-field SDE**

Promisingly, in addition to the fact that the SDE efficiency and polarity can be switched simply by electrostatic gating $V_{PG}$, we are also able to actively control the maximal efficiency of the zero-field SDE via $R_L$ and $I_C$ respectively (Fig. 4) after understanding the origin of the SDE as $|eI_C R_L|$. In the $R_L$-tuning approach, Figs. 4a-c shows the typical $I_{C+,-}(V_{PG})$, $\Delta I_C(V_{PG})$ and the diode efficiency $\eta(V_{PG}) = 2\Delta I_C/(I_{C+}+|I_{C-}|)$ by increasing room-temperature resistance $R_L = 6$ kΩ, 11 kΩ, and 16 kΩ, while keeping $I_C \approx 48$ nA. In this approach, both the relative shift in $V_{PG}$ between $I_{C+}$ and $|I_{C-}|$ (defined as d$V_{PG}$ in top panel of Fig. 4c) and the maximal absolute efficiency $|\eta_{max}|$ (defined in bottom panel of Fig. 4c) increases with $R_L$ (and thus $|I_C R_L|$). Alternatively, in the $I_C$-tuning approach, we fix the high $R_L = 101$ kΩ and vary $I_C$ from ~ 9 nA to 36 nA by the paralleled JJ controlled via $V_{SG}$ in Figs. 4d-f. Here d$V_{PG}$ can be made even larger, reaching over half of the $V_{PG}$ period in Figs. 4e, f, whereas $|\eta_{max}|$ decreases with $I_C$ (and thus $|I_C R_L|$) in this approach. The reason for the latter is because here the maximal $|\Delta I_{C,max}|$ (denoted in Fig. 4d, mid panel) is relatively stable and almost saturates with $I_C$ once the shift d$V_{PG}$ reaches half of the $V_{PG}$ period while $I_C$ keeps growing. Fig. 4g, h summarize all d$V_{PG}$ and $|\eta_{max}|$ versus $|I_C R_L|$ in the $R_L$-tuning and $I_C$-tuning approaches, respectively (The original dataset is included in Supplementary Information Sec. 5). For each point $|I_C R_L|$ takes the averaged $I_{C+}$ over the whole period. The increasing d$V_{PG}$ with increasing $|I_C R_L|$ further confirms the chemical potential shifting as the origin for SDE here. A highest $|\eta_{max}|$ ~ 60% is achieved (highlighted by the red dashed circle in Fig. 4h, lower panel), which can in principle be further increased by optimizing the device design. In each configuration, $V_{PG}$ can be used as knob to turn on/off the SDE and switch the diode polarity, further enhancing its controllability.

We note that the voltage shift of the island by $I_C R_L$ causes a relative shift in the gating effect of $V_{PG}$. This effect also shifts $I_{C+,-}(V_{PG})$ in the same direction as the chemical potential shift mechanism, but is negligible here since $I_C R_L$ is much smaller than the observed d$V_{PG}$ (Figs. 4g, h). We also note that the strongly tilted $I_{C+,-}(V_{PG})$ observed in Figs. 4d-f, which is different from the non-tilted $I_{C+,-}(V_{PG})$ in the CPT without SDE[36-38], can also be qualitatively explained in the framework of chemical potential shift (see Supplementary Fig. 9d). In addition, $|\eta_{max}|$ may have difference values with similar $|I_C R_L|$ (such as $|\eta_{max}| \approx 8\%$ at $|I_C R_L| \approx 1.0$ mV in Figs. 4g (lower panel) while $|\eta_{max}| \approx 60\%$ at $|I_C R_L| \approx 1.0$ mV in Figs. 4h (lower panel)). This is because the two cases has similar $|\Delta I_{C,max}|$ but different different $I_C$ from the paralleled JJ.

**Discussion on relation with existing designs**

Several works utilize magnetic vortices and Meissner effect to create SDE in classical superconducting thin-films[45,46]. However, such proposals usually require magnetic field to operate. Also, due to the unstable and uncontrollable nature of the vortices[47], the reliability of such scheme is limited and may be highly dependent on film quality, especially when extending the system into more complex architectures. In comparison, the zero-field SDE here guarantees a high controllability, since it is entirely independent of the intrinsic ISB/TRSB inside the superconductor and is determined by external circuitry. In addition, by choosing the superconducting device with well-understood materials and mature fabrication techniques such as CPT here, a high yield of the SDE can be readily achieved, which facilitates further extensible engineering. The generic SDE design here can also be easily extended to high-$T_C$ superconductors, which enhances the operational temperature. Moreover, the on-off states of the SDE as well as its polarity can be tuned simply by electric gating. Such features resemble the field-effect transistors (FETs) in semiconducting electronics, which is notably difficult to achieve under the prevailing superconducting electronics technologies based on the superconducting quantum flux (SFQ) in conventional Josephson junctions[48]. This may significantly enhance the flexibility and scalability of the future superconducting electronics based on the SDE design here, particularly because the logic operations can be realized by voltage levels as in FETs instead of pulses in SFQ devices. Also, the gate-controlled SDE, which in principle can also be operated as the superconducting triode, with its voltage level-based logic further facilitates cascading multiple SDEs for advanced functionalities.

We emphasize that the zero-field SDE without material-specific TRSB/ISB here does not contradict the general relation between SDE and TRSB/ISB[4-9], but only expands the Hamiltonian of the SDE with only the superconducting system to include its external circuitry, where TRSB/ISB occurs in the finite voltage drop in $R_L$ besides intrinsically in the superconducting system. On the other hand, our work suggests alternative scenarios in interpreting zero-field SDE in the search for exotic and intrinsic ISB/TRSB superconductors[24-29,49,50]. Indeed, the chemical potential shift, while negligible for metallic superconductors, may become significant for high-$T_C$ superconductors, low-dimensional materials and hybrid devices[31-35,51]. Our work thus expands the current understanding of the relationship between SDE and TRSB in superconductivity and highlights the underappreciated but crucial role of external circuitry in constructing the SDE.

**Conclusions**

In conclusion, using the CPT based on Al-InAs nanowire as a prototype, we demonstrated a new type of zero-field SDE due to the generic chemical potential shift in the external circuit, which is applicable to wide range of materials/devices without relying on the material-specific ISB/TRSB mechanisms of the superconductor itself. Its polarity and efficiency can be modified by easy control of $V_{PG}$ as well as by biasing circuitry. In addition, the optimization and extended engineering can in principle be largely done via $R_L$ at room temperature, avoiding the stringent space and heat-load limitations imposed by low-temperature environment. Such generic, controllable and extensible features, which highlight a paradigm shift from material-limited to circuit-level-engineered design, may greatly facilitate the future development of using superconducting diodes to construct complex superconducting electronics towards ultra-low dissipation and noise electronic systems, bridging the gap between quantum device physics and practical circuit engineering.

**Method**

*InAs-Al Nanowire Growth*: All the InAs nanowires were grown in a solid source molecular-beam epitaxy (MBE) system (VG V80H) on the commercial n-type Si (111) substrates using Ag as catalysts. Before loading the Si substrates into the MBE chamber, they were pretreated in a diluted HF (2%) solution for 1 min to remove the surface contamination and native oxide[41]. For Ag catalyst deposition, an Ag layer less than 0.5 nm nominal thickness was deposited on the Si(111) substrate in the MBE growth chamber at room temperature and then annealed *in situ* at 550 °C for 20 min to generate small Ag nanoparticles. Ultrathin InAs nanowires with a diameter ranging from ~20 to 40 nm were obtained with these small Ag nanoparticles. The ultra-thin InAs nanowires were grown for 80 min at a temperature of 485 °C with an arsenic/indium beam equivalent pressure ratio of ~ 42 (the beam fluxes of In and As$_4$ sources are $1.1\times10^{-7}$ mbar and $4.6\times10^{-6}$ mbar, respectively). After the growth of the InAs nanowires, the sample was transferred from the growth chamber to the preparation chamber at 300 °C to avoid arsenic condensation on the nanowire surface. The sample was then cooled down to a low temperature (~-40 °C) by natural cooling and liquid nitrogen cooling[42]. Al was evaporated from a Knudsen cell at an angle of ~20° from the substrate normal (~70° from substrate surface) and at a temperature of ~1150 °C for 100 s (giving approximately 0.08 nm/s). To obtain half Al shells, the substrate rotation was kept disabled during the Al growth. When the growth of InAs-Al nanowires was completed, the sample was rapidly pulled out of the MBE growth chamber and oxidized naturally.

*Device fabrication*: InAs-Al nanowires were transferred by a wiper. The wiper first gently swiped the growth substrate and then swiped again on the device substrate which was highly p-doped Si covered by 300 nm silicon dioxide. The randomly deposited nanowires are then selected via scanning electron microscopy (SEM) with minimal exposure time. With standard electron beam lithography processes, the contact areas were defined. Tunnel barriers at both ends of the nanowire formed by etching the aluminum (Al) thin film using Transene Aluminum Etchant Type D at 50°C for 10 seconds. Ohmic contacts to the InAs nanowire were fabricated by 80 s Ar plasma etching at a power of 50 W and pressure of 0.05 Torr, followed by metal deposition of Ti/Al (5/65 nm) bilayer.

*Measurement technique*: The two types of electrical measurement techniques (lock-in amplifier measurement technique and fast counter measurement technique) are explained in details in Supplementary Information Sec. 1.

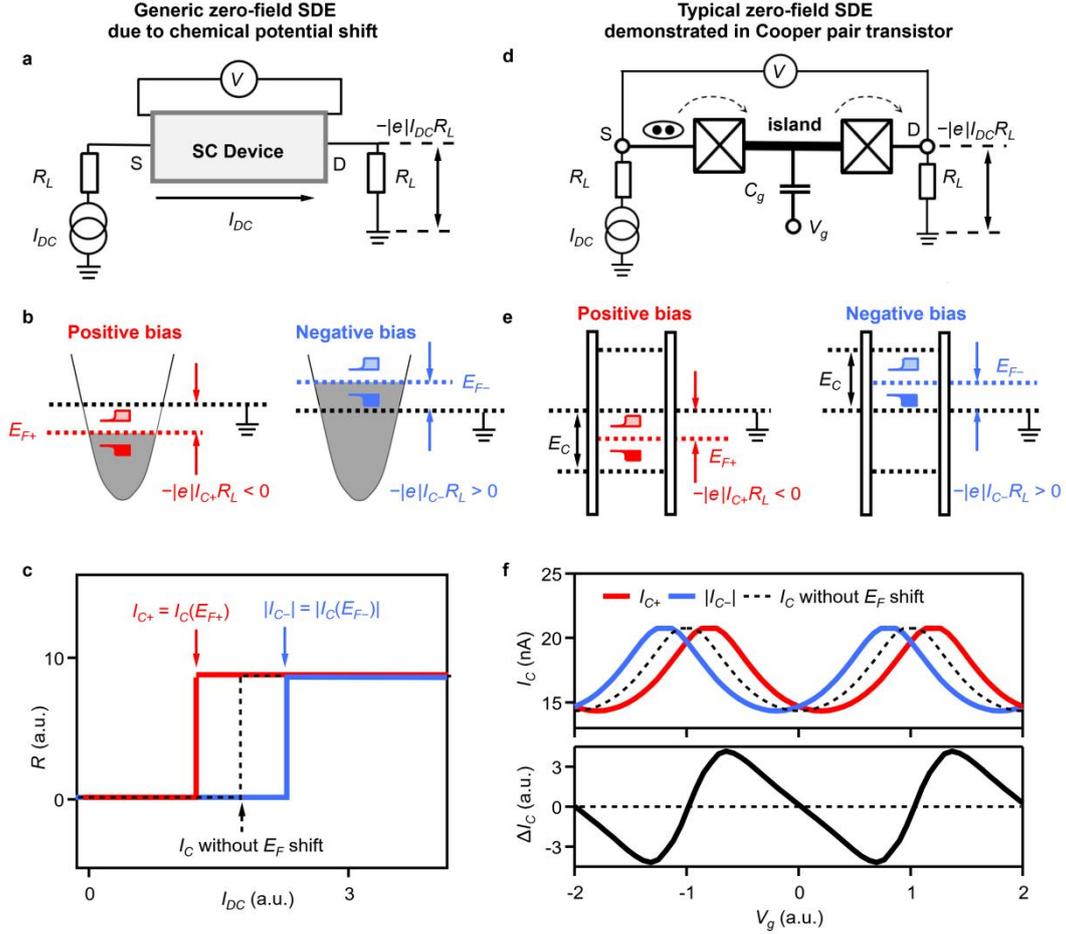

**Fig. 1. Principle of zero-field SDE caused by line resistance-induced chemical potential shift without ISB and TRSB**: **a,** Schematic of the generic circuitry for measuring SDE. The superconducting device at low temperature has S and D contacts. D is connected to ground via the generic $R_L$. $I_{DC}$ shift the chemical potential of the device by $-|e|R_L I_{DC}$. The other $R_L$ connecting S in **a**, **d** does not affect the device chemical potential. **b,** $E_{F+,-}$ of the superconducting device is shifted down/up (red/blue lines) with positive $I_{C+}$/negative $I_{C-}$. The superconducting gap opened at $E_{F+,-}$ is drawn. **c,** Schematic $R(I_{DC})$ showing SDE due to $E_F$ shift (see text). $R(I_{DC})$ without such shift is marked by the black dashed line. Such zero-field SDE induced by chemical potential shift is also demonstrated in CPT in **d-f**. **d,** Schematic of CPT. The island transmitting single Cooper-pair is marked by thicker line, capacitively coupled to gate and tunnel-coupled to S and D via JJs (crossed squares). **e,** $E_F$ shift of the

island due to similar mechanism in **b**. $I_C$ is sensitive to $E_F$ shift comparable to $E_C$ due to Coulomb blockade effect. **f,** $I_{C+}(V_{PG})$ (red) and $|I_{C-}(V_{PG})|$ (blue) each oscillate $V_{PG}$ but shifted right/left due to $E_F$ shift (top panel, see text), causing a gate-dependent SDE ($\Delta I_C = I_{C+} - |I_{C-}|$ in the bottom panel). $I_C(V_{PG})$ without such shift is marked by the black dashed lines (top panel).

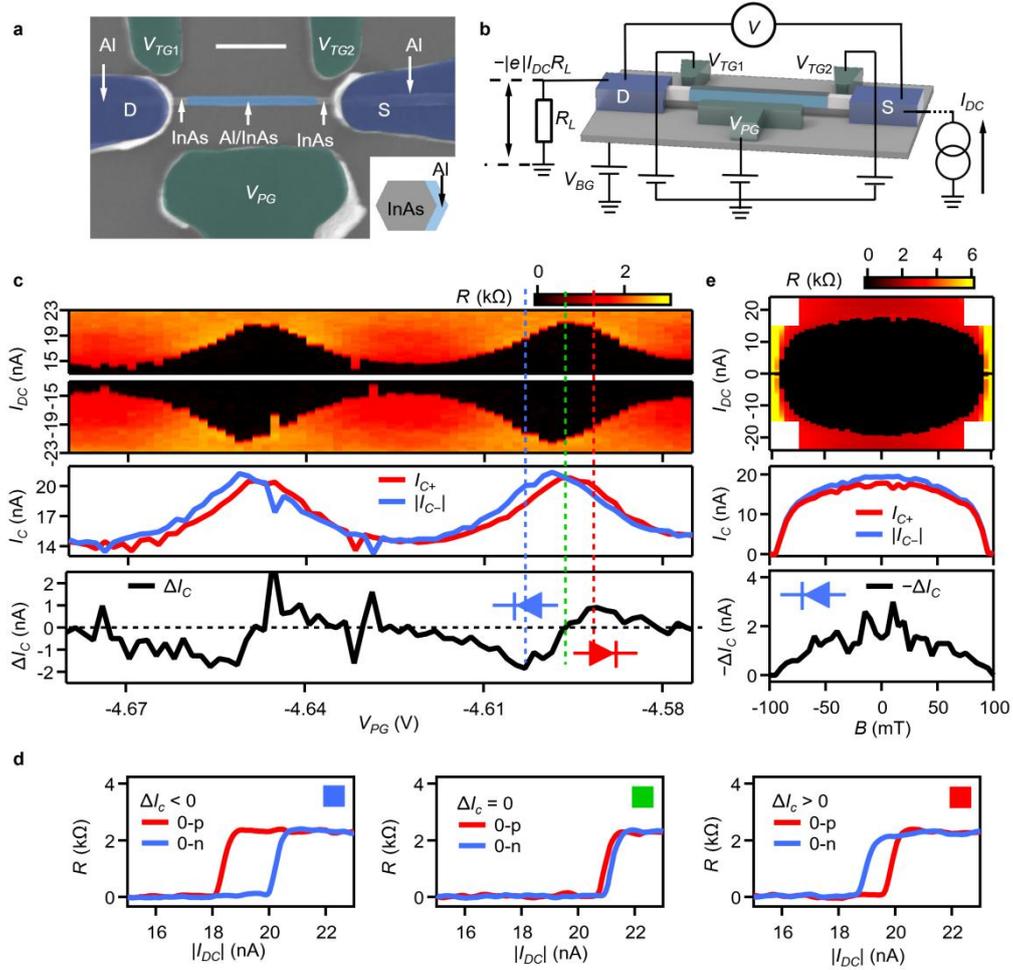

**Fig. 2. Gate-switchable and zero-field SDE in CPT (D1): a,** False-color SEM image of the device. (inset) cross-section of the epitaxial Al on InAs nanowire. Scale bar: 500 nm. The island with epitaxial Al (light blue) is tunnel-coupled to Al S/D (dark blue) via a short section of bare InAs (light gray). **b,** Schematic for the setup. The dotted line in the circuit denoted the other $R_L$ not affecting the device chemical potential. **c,** (Top panel) lock-in measurement of $R$ = d$V$/d$I$ versus $I_{DC}$ and $V_{PG}$ (top panel). ($V_{TG1}$, $V_{TG2}$, $V_{BG}$) = (0 V, 0 V, 0 V). $I_C$ oscillates $V_{PG}$ due to Coulomb blockade effects (see text). $T \approx 9$ mK, $B = 0$ mT. The extracted $I_{C+}(V_{PG})$ is shifted to higher $V_{PG}$ relative to $|I_{C-}|(V_{PG})$ (middle panel), resulting in $\Delta I_C = I_{C+} - |I_{C-}|$ oscillating with $V_{PG}$ (bottom panel) and showing clear gate-switchable SDE (Fig. 1f) with negative polarity (blue dashed line), positive polarity (red dashed line), and no SDE (green dashed line). **d,** Linecuts in **c**: $R(I_{DC})$ along the dashed lines of the same color in **c. e,** $R(I_{DC}, B)$

at $V_{PG}$ for the blue dashed line in **c** (top panel). $I_{C+}$ and $|I_{C-}|$ (middle panel) and $-\Delta I_C(B)$ (bottom panel). $\Delta I_C(B)$ is an even function (the polarity unchanged by field) and maximal at $B$ = 0 mT, confirming SDE is zero-field (see text). $B$ in **e** is paralleled with the nanowire. See Supplementary Fig. S5 for data of $B$ in other directions.

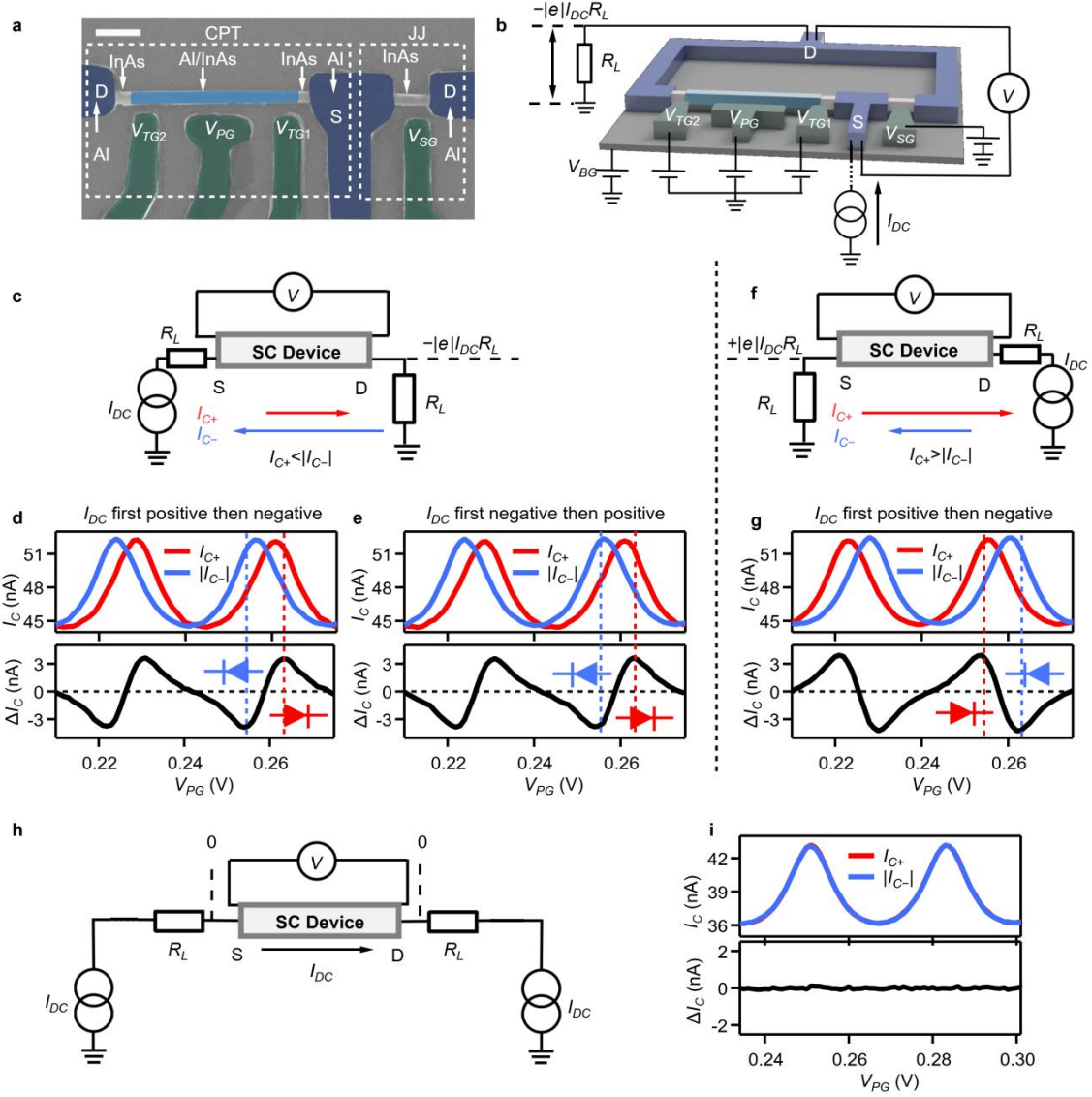

**Fig. 3. Zero-field SDE in D3 determined by chemical potential shift in external circuit not by intrinsic ISB/TRSB of superconducting device**: **a,** SEM of D3. Scale bar: 200 nm. **b,** Schematic of the setup. The CPT is in parallel with another JJ for additional $I_C$ tuning via $V_{SG}$. The dotted line in the circuit denoted the other $R_L$ not affecting the device chemical potential. **c,** Schematic with D grounded and -|e|$I_{DC}R_L$. **d,** $I_{C+}(V_{PG})$, |$I_{C-}(V_{PG})$| and $\Delta I_C(V_{PG})$ measured with **c,** by the fast counter technique (see text), reproducing similar zero-field SDE in Fig. 2c. $I_{C+}$ (red arrow) is smaller than |$I_{C+}$| (blue arrow) at the blue dashed line, illustrated in **c**. $I_{DC}$ first

sweeps to positive then to negative direction for each $V_{PG}$. **e,** Similar measurement with the opposite sweep order, producing the same SDE polarity. **f,** Schematic with S grounded and $+|e|I_{DC}R_L$. **g,** Similar measurement with **f**, producing reversed SDE polarity. For **d-f**: ($V_{TG1}$, $V_{TG2}$, $V_{BG}$, $V_{SG}$) = (-1.2 V, -2.4 V, -3.5 V, 0.0 V). Such polarity reversal (blue/red arrows in **f**) contradicts the intrinsic ISB/TRSB of the CPT. **h,** Schematic of "symmetric biasing" with $|e|I_{DC}R_L = 0$ (see text). **i,** Similar measurement showing no SDE. ($V_{TG1}$, $V_{TG2}$, $V_{BG}$, $V_{SG}$) = (-1.2 V, -2.4 V, -3.5 V, 1.0 V). $T \approx 8$ mK, $B = 0$ mT. (D, E, G): $R_L = 21$ k$\Omega$, (I): $R_L = 11$ k$\Omega$.

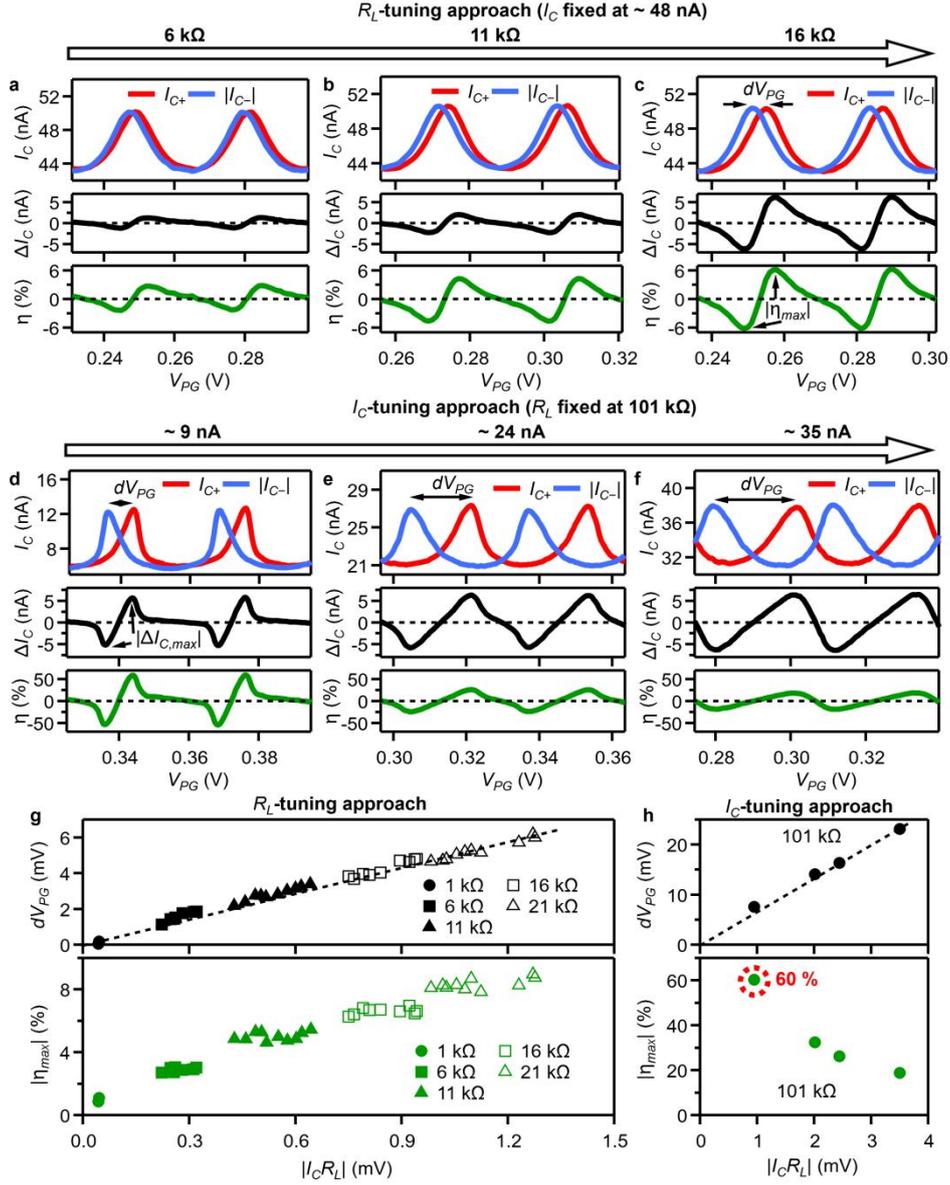

**Fig. 4. Zero-field SDE efficiency of D3 controlled by $R_L$ and $I_C$: a-c,** $I_{C+,-}(V_{PG})$ (red, blue), $\Delta I_C(V_{PG})$ (black) and SDE efficiency $\eta(V_{PG}) = 2\Delta I_C/(I_{C+}+|I_{C-}|)$ (green) measured with similar $I_C$ but different $R_L = 6$ kΩ, 11 kΩ, 16 kΩ, respectively. $dV_{PG}$ and $|\eta_{max}|$ are defined in **c**. **a**: $(V_{TG1}, V_{TG2}, V_{BG}, V_{SG}) = (-1.2$ V, -2.4 V, -3.5 V, 1.5 V). **b**: $(V_{TG1}, V_{TG2}, V_{BG}, V_{SG}) = (-1.2$ V, -2.4 V, -3.5 V, 0.0 V). **c**: $(V_{TG1}, V_{TG2}, V_{BG}, V_{SG}) = (-1.2$ V, -2.4 V, -3.5 V, 0.5 V). **d-f,** Same as **a-c** but with fixed $R_L = 101$ kΩ and varied $I_C$ by $V_{SG}$ of the paralleled JJ (see Figs. 3a, b). $dV_{PG}$ is larger than the half period of $V_{PG}$ in **e, f** with large $|I_C R_L|$. **d**: $(V_{TG1}, V_{TG2}, V_{BG}, V_{SG}) = (4.5$ V,

1.4 V, -13.8 V, -3.0 V). **e:** ($V_{TG1}$, $V_{TG2}$, $V_{BG}$, $V_{SG}$) = (4.5 V, 1.4 V, -13.8 V, 2.9 V). **f:** ($V_{TG1}$, $V_{TG2}$, $V_{BG}$, $V_{SG}$) = (-1.2 V, -2.5 V, -3.6 V, -3.3 V). **g, h,** Summary of all d$V_{PG}$ (upper panel) and |$\eta_{max}$| (lower panel) versus |$I_C R_L$| for the $R_L$-tuning approach **g** and the $I_C$-tuning approach **h**. d$V_{PG}$ versus |$I_C R_L$| are extrapolated to origin with (d$V_{PG}$, |$I_C R_L$|) = (0 V, 0V) (black dashed lines, top panels). The maximal |$\eta_{max}$| ≈ 60 % (corresponding to **d**, red dashed circle). $T$ ≈ 8 mK, $B$ = 0 mT.

## Data availability

All data needed to evaluate the conclusions in the paper are present in the main text and/or the supplementary information. Raw data generated in this study are available from the corresponding author upon reasonable request.

## Acknowledgement

We are grateful to Cong-jun Wu, Jian Li, Da Wang, Kun Jiang, Chunxiao Liu, Leo Kouwenhoven, Constantin Schrade for helpful discussions. The work of Z.D. and J.S. was supported by the Young Scientists Fund of the National Natural Science Foundation of China (Grant No. 2024YFA1613200) The work of J.H. was supported by the National Key Research and Development Program of China (Grant No. 2022YFA1403901) and the new cornerstone investigator program. The work of D.P. and J.Z. was supported by the National Natural Science Foundation of China (Grant Nos. 12374459, 61974138 and 92065106), the Innovation Program for Quantum Science and Technology (Grant 2021ZD0302400). D. P. acknowledges the support from Youth Innovation Promotion Association, Chinese Academy of Sciences (Nos. 2017156 and Y2021043). The work of J.S., L.L., F.Q. and G.L. were supported by the National Key Research and Development Program of China (Grant Nos. 2023YFA1607400), the Beijing Natural Science Foundation (Grant No. JQ23022), the Strategic Priority Research Program B of Chinese Academy of Sciences (Grant No. XDB33000000), the National Natural Science Foundation of China (Grant Nos. 92065203, 12174430, 92365302), and the Synergetic Extreme Condition User Facility (SECUF, https://cstr.cn/31123.02.SECUF). Y.L. acknowledges support from National Natural

Science Foundation of China, Grant No. 12404154. The work of other authors were supported by the National Key Research and Development Program of China (Grant Nos. 2019YFA0308000, 2022YFA1403800, 2023YFA1406500, and 2024YFA1408400), the National Natural Science Foundation of China (Grant Nos. 12274436, 12274459), the Beijing Natural Science Foundation (Grant No. Z200005), and the Synergetic Extreme Condition User Facility (SECUF, https://cstr.cn/31123.02.SECUF). The work is also funded by Chinese Academy of Sciences President's International Fellowship Initiative (Grant No. 2024PG0003).

**Author contributions**

J.S., J.P.H. L.L. conceived and designed the experiment.

S.F., G.L. fabricated the devices and performed the transport measurements, with assistance from Y. L., A. W., Z. Z., X. G., X. D., discussed with B.T., Z.L., P.L., F.Q., G.L, supervised by Z.D., L.L., J.S.

D.P. and J.Z. grew nanowire materials.

S.F., Z.D. performed the theoretical analysis, discussed with J.S., J.P.H..

S.F., Z.D., G.L., J.S. analyzed the data and wrote the manuscript, with input from all authors.

**Competing interests**

Authors declare that they have no competing interests.

Phys. Rev. Lett. 131, 027001 (2023).

# Supplemental Information: Circuit-level-configurable Zero-field Superconducting Diodes: A Universal Platform Beyond Intrinsic Symmetry Breaking

Xiaofan Shi[1,4,†], Ziwei Dou[1,†,*], Dong Pan[2,†], Guoan Li[1,4,†], Yupeng Li[1,5], Anqi Wang[1],

Zhiyuan Zhang[1,4], Xingchen Guo[1,4], Xiao Deng[1,4], Bingbing Tong[1], Zhaozheng Lyu[1], Peiling

Li[1], Fanming Qu[1,4,6,7], Guangtong Liu[1,6,7], Jianhua Zhao[2,3*], Jiangping Hu[1,8,9*], Li Lu[1,4,6,7*],

Jie Shen[1,6*]

[1]Beijing National Laboratory for Condensed Matter Physics and Institute of Physics, Chinese

Academy of Sciences, Beijing 100190, China

[2]State Key Laboratory of Semiconductor Physics and Chip Technologies, Institute of

Semiconductors, Chinese Academy of Sciences, Beijing 100083, China

[3]National Key Laboratory of Spintronics, Hangzhou International Innovation Institute,

Beihang University, Hangzhou 311115, China

[4]University of Chinese Academy of Sciences, Beijing 100049, China

[5]Hangzhou Key Laboratory of Quantum Matter, School of Physics, Hangzhou Normal

University, Hangzhou 311121, China

[6]Songshan Lake Materials Laboratory, Dongguan, Guangdong 523808, China

[7]Hefei National Laboratory, Hefei 230088, China

[8]Kavli Institute of Theoretical Sciences, University of Chinese Academy of Sciences, Beijing,

100190, China

[9]New Cornerstone Science Laboratory, Shenzhen 518054, China

[*] Corresponding authors. E-mails: ziweidou@iphy.ac.cn, jhzhao@semi.ac.cn,

jphu@iphy.ac.cn, lilu@iphy.ac.cn, shenjie@iphy.ac.cn,

[†]These authors contributed equally to this work.

Table of Content

# 1 Lock-in amplifier and fast counter measurement setups for $I_{C+,-}$

Fig. 2 uses the conventional lock-in amplifier setup whose detailed circuit is illustrated in Supplementary Fig. 1a. Besides the resistance $r$ already installed inside the dilution refrigerator[1], another $R$ at room temperature is inserted for direct measurement of ac current response through the device. The dc current bias $I_{DC}$ is provided by a high-precision digital source. The ac current $I_{AC}$ is provided by the lock-in amplifier and is added to $I_{DC}$. The total line resistance shifting the chemical potential of the device is $R_L = r+R$ (red dashed box). The ac voltage across the device and the ac current through the device are thus demodulated by the lock-in amplifier, whose in-phase component are $V_{AC}$ and $I_{AC}$, respectively. The frequency of $I_{AC}$ is chosen such that phase angle is close to 0. The differential resistance of the device is thus $dV/dI$. The lock-in measurement is usually slow since the demodulation time should be several times longer than the period of $I_{AC}$ and the total measurement time of $I_{C+,-}$ for a given $V_{PG}$ multiplies with the number of points in $I_{DC}$.

The fast counter measurement can obtain $I_{C+,-}$ in a more efficient way, and is widely accepted in other Josephson junction and SDE measurements[2-4]. In such setup, a triangular waveform $I_{DC}$ is produced by a digital signal generator, and the dc voltage of the device $V_{+,-}$ is amplified and filtered to optimize its signal-to-noise ratio before sending to the digital counter. Meanwhile, a square wave (TTL) synchronized with $I_{DC}$ is also sent to the counter. The counter is configured to register the time at the rising edge of the input waveforms. Therefore, the time at $I_{DC} = 0$ is registered by the rising edge of the square wave, and the time at $I_{DC} = I_{C+,-}$ is registered when $V_{+,-}$ jumps abruptly from superconducting (zero-voltage) and finite-voltage state. To enhance the accuracy of $I_{C+,-}$, the average from several repeated measurement is done at each $V_{PG}$. The measurement time of $I_{C+,-}$ for a given $V_{PG}$ is set by the period $T$ of $I_{DC}$ multiplied by the number of repeated measurement, which makes it much faster than the lock-in technique. $T$ is chosen such that the results in Figs. 3 and 4 are independent of its period.

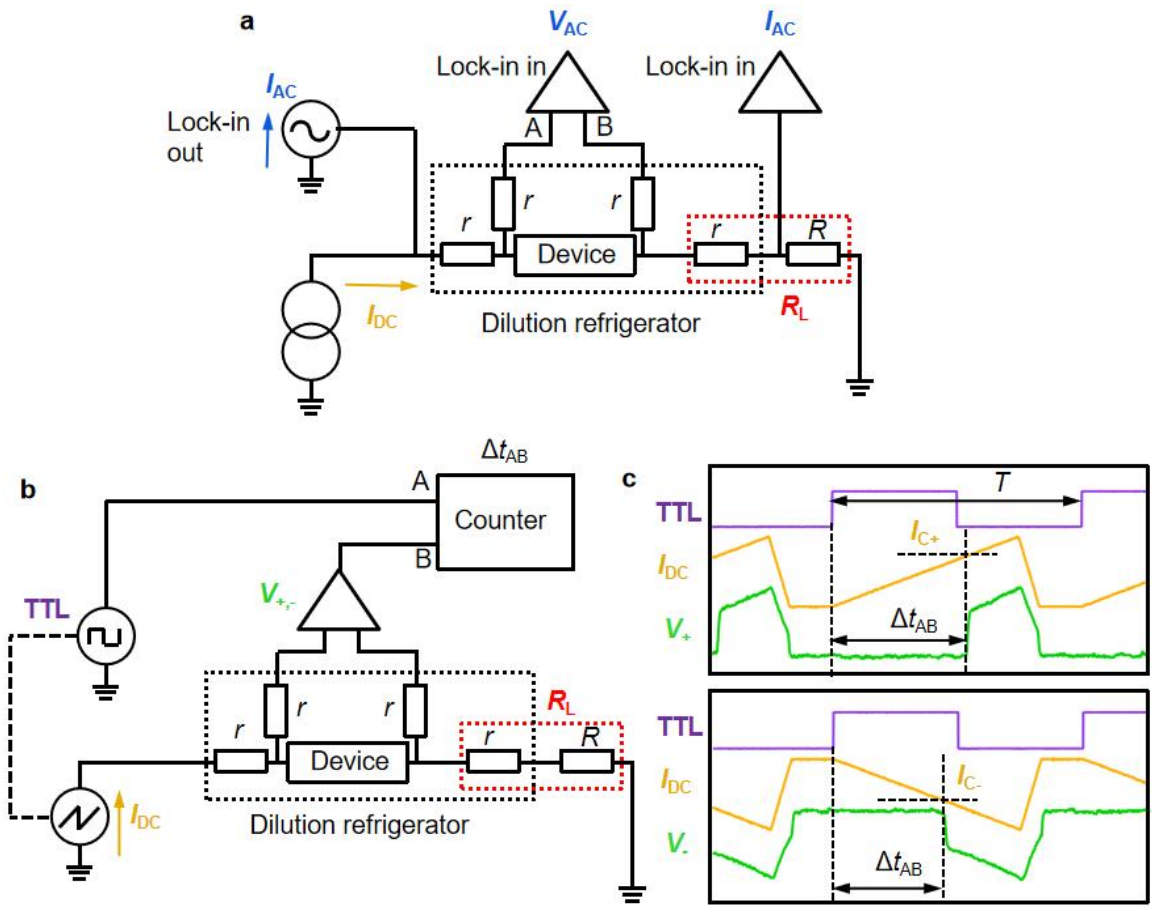

**Supplementary Fig. 1. Schematic for lock-in amplifier and fast counter measurement setups: a,** Lock-in amplifier measurement setup: Black dashed box: Device and line resistance $r = 1$ k$\Omega$ inside the dilution refrigerator. Another $R$ at room temperature is grounded and can be changed during the measurement. The total line-resistance in the main text is thus $R_L = r + R$ (red dashed box). DC source and "lock-in out" provide $I_{DC}$ (yellow) and the additional small $I_{AC}$ (blue) to the device, respectively. $V_{AC}$ and $I_{AC}$ (blue) demodulated by lock-in amplifiers are used to measure the device differential resistance $V_{AC}/I_{AC}$. **b,** Fast counter setup: a signal generator produces a triangular wave $I_{DC}(t)$ (yellow waveform in **c**) with period $T$. The device voltage $V_{+,-}$ (green waveform in **c**) is 0 below $I_{C+,-}$ and has a sharp jump at $I_{C+,-}$. $I_{C+,-}$ are measured by the time difference $\Delta t_{AB}$ registered in a digital counter, between the jump and $I_{DC} = 0$, set by a synchronized square wave (TTL, purple waveform in **c**). **c,** An example of the time-domain waveforms in positive (upper) and negative (lower) directions.

## 2 SDE in critical and retrap current

Supplementary Fig. 2a shows the typical differential resistance in D1 $R = \mathrm{d}V/\mathrm{d}I$ versus $I_{DC}$ (top panel) and the absolute dc voltage across the device $|V|$ versus $I_{DC}$ (bottom panel). $|V|$ is obtained by numerically integrated the measured $R(I_{DC})$. Same as all the data in the main text and the other sections in Supplementary Information, $I_{DC}$ in Supplementary Fig. 2a is swept from 0 to either positive direction (marked as "0-p") or negative direction ("0-n"). The current at which the resistance becomes finite corresponds to the critical current, noted as $I_{C+,-}$. At suitable gate and $R_L$ configuration, $\Delta I_C = I_{C+} - |I_{C-}|$ and the SDE appears. On the other hand, Supplementary Fig. 2b shows the similar $R(I_{DC})$ (top panel) and $|V|(I_{DC})$ (bottom panel) at the same configurations as Supplementary Fig. 2a, but with $I_{DC}$ swept from finite resistance state to superconducting state ("p-0" or "n-0" in Supplementary Fig. 2b). The current at which $R$ returns from finite value to 0 is noted as the "retrap current"[5] $I_{R+,-}$. Here another SDE[6,7] defined as $\Delta I_R = I_{R+} - |I_{R-}|$ is observed. Different from $\Delta I_C$ always starting from the zero-voltage state, such $\Delta I_R$ can be influenced by the finite-voltage state which already breaks TRSB without magnetic field, ferromagnetism, or other spontaneous polarization mechanism[8-12], and is similar to the semiconducting diode[6,7]. We emphasize that all our measurement in the main text are the SDE in $\Delta I_C$, not $\Delta I_R$.

We also note that $I_{C+} \neq I_{R+}$ and $I_{C-} \neq I_{R-}$ in Supplementary Figs. 2a, b. This difference is present in Josephson junctions with high normal resistance and geometric capacitance between S/D electrodes (that is, low damping)[5]. Such different $I_C$ and $I_R$ can also happen in junctions with low normal resistance and capacitance, possibly due to the heating effect in the finite-voltage state reducing $I_R$ below $I_C$[13]. The detailed mechanism does not affect the interpretation of $\Delta I_C$ as caused by the chemical potential shift in our devices in the main text.

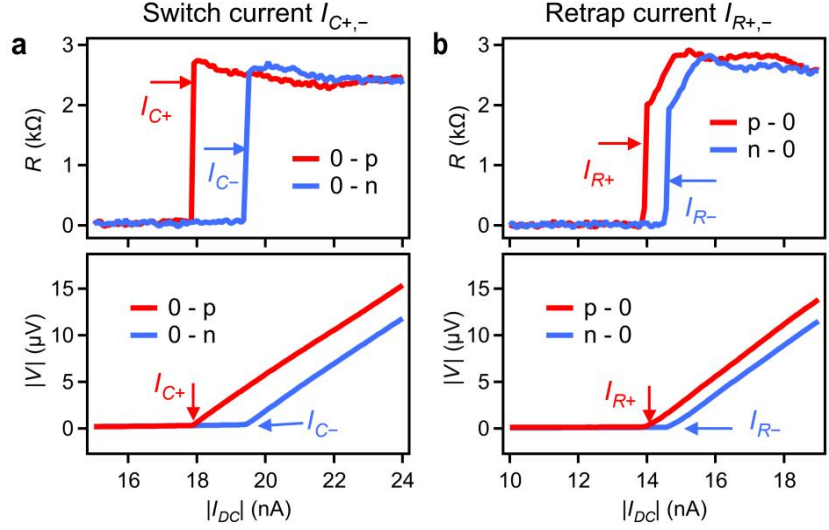

**Supplementary Fig. 2. SDE in critical and retrap current (D1). a,** Top panel: $R = \mathrm{d}V/\mathrm{d}I$ versus $|I_{\mathrm{DC}}|$ showing clear SDE in the critical current $\Delta I_{\mathrm{C}} = I_{\mathrm{C+}} - |I_{\mathrm{C-}}|$. $|I_{\mathrm{DC}}|$ is swept from 0 to above $|I_{\mathrm{C+,-}}|$. Bottom panel: the absolute dc voltage across the device $|V|$ versus $|I_{\mathrm{DC}}|$, obtained from integrating the measured $R(I_{\mathrm{DC}})$ in the top panel data. **b,** Similar to **a** but with $|I_{\mathrm{DC}}|$ swept from above $|I_{\mathrm{C+,-}}|$ to 0, showing another SDE in the retrap current $\Delta I_{\mathrm{R}}$ in the same gate configuration as **a**. $R_{\mathrm{L}} = 11$ k$\Omega$. $T \approx 10$ mK and $B = 0$.

## 3 Zero-field SDE in D1 with opposite sweep order of $I_{DC}$, S/D as grounded contact, and opposite sweep direction of $V_{PG}$

Similar to D3 in Fig. 3, we also provide similar data for the D1 shown in Fig. 2, measured with lock-in amplifiers (see Fig. 2b and Supplementary Fig. 1a). The device is grounded via a total $R_L = 11$ k$\Omega$. The CPT is first grounded by D via $R_L = 11$ k$\Omega$ with the positive $I_{DC}$ direction defined as the arrow (from S to D, Supplementary Fig. 3a,b). Supplementary Fig. 3c displays similar SDE dependence on $V_{PG}$, measured with opposite order of $I_{DC}$ sweep. The $\Delta I_C(V_{PG})$ is reproduced, independent of $I_{DC}$ sweep order, same as Figs. 3d, e.

The CPT is then grounded by S via $R_L = 11$ k$\Omega$ with the positive $I_{DC}$ direction still defined as from S to D (Supplementary Fig. 3d). Supplementary Fig. 3e shows reversed SDE polarity by swapping S and D to be connected to ground, also reproducing Fig. 3g, confirming that the zero-field SDE is not caused by intrinsic ISB/TRSB of the CPT which is not changed by swapping the grounded contacts.

We also show similar SDE with D connected to ground (same as Supplementary Fig. 3a), while $V_{PG}$ steps increasingly (Supplementary Fig. 3f) or decreasingly (Supplementary Fig. 3g). The non-hysterical behavior to gate proves that the SDE is not caused by spontaneously polarized spin/valley at certain Fermi level[8,9].

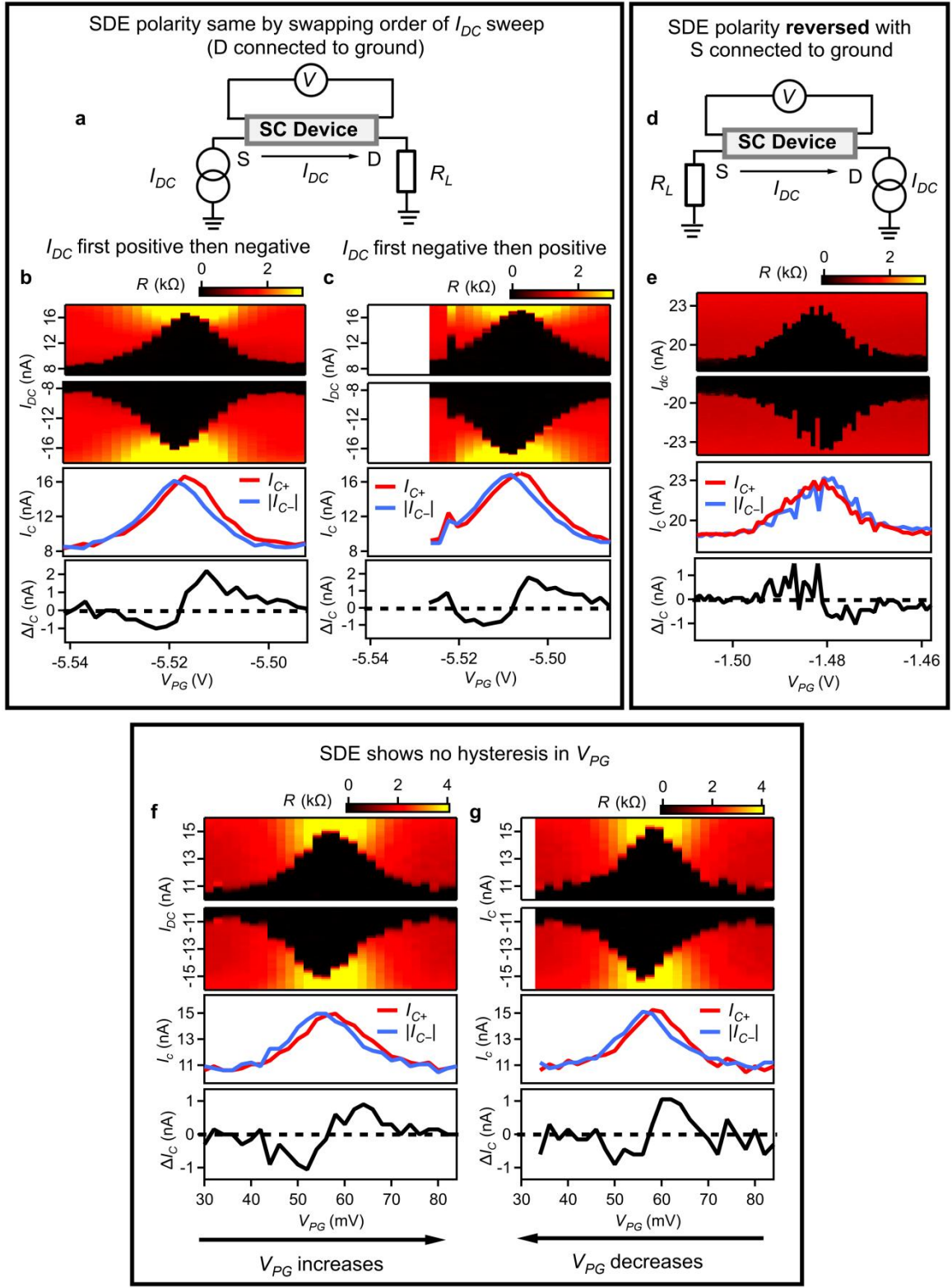

**Supplementary Fig. 3. More data for zero-field SDE in D1 with opposite sweep order of**

$I_{DC}$**, S/D as grounded contact, and opposite sweep direction of** $V_{PG}$**: a,** Schematic circuit

with D connected ground via $R_L$ = 11 kΩ. Positive $I_{DC}$ is defined as from S to D. **b,** $R(V_{PG}, I_{DC})$

(top panel), the extracted $I_{C+,-}(V_{PG})$ (middle panel) and $\Delta I_C(V_{PG})$ (bottom panel). The sweep order of $I_{DC}$: first 0 to positive $I_{DC} > I_{C+}$, then 0 to negative $I_{DC} < -|I_{C-}|$ for each $V_{PG}$. **c,** Similar to **b** but with the opposite sweep order: The sweep order of $I_{DC}$: first 0 to negative $I_{DC} < -|I_{C-}|$, then 0 to positive $I_{DC} > I_{C+}$ for each $V_{PG}$. Gate configuration for **b**, **c**: $V_{TG1} = 0$ V, $V_{TG2} = -5.9$ V, $V_{BG} = 0$ V. **d,** Schematic circuit with S connected ground via $R_L = 11$ kΩ. Positive $I_{DC}$ is defined as from S to D. **e,** Similar to **b** but with setup in **d**. The sweep order of $I_{DC}$: first 0 to positive $I_{DC} > I_{C+}$, then 0 to negative $I_{DC} < -|I_{C-}|$ for each $V_{PG}$. $V_{TG1} = 0$ V, $V_{TG1} = 0$ V, $V_{BG} = 0$ V. **f, g,** Similar to **b** (same setup and sweep order of $I_{DC}$). $V_{TG1} = -3$ V, $V_{TG1} = -3$ V, $V_{BG} = 0$. with increasing $V_{PG}$ (**f**) and decreasing $V_{PG}$ (**g**). All measurements are done at $T \approx 10$ mK and $B = 0$.

**4 Zero-field SDE reproduced in D2 and all B-dependent $\Delta I_C$ data for D1 and D2**

The zero-field SDE is reproduced in another nominally identical device (D2), whose SEM image is shown in Supplementary Fig. 4a. The device is measured with the same lock-in setup as Fig. 2, and $R_L$ = 11 k$\Omega$ is connected to D. Supplementary Fig. 4b shows $R(V_{PG}, I_{DC})$ under positive and negative sweeps of $I_{DC}$. Similar to Fig. 2c, without magnetic field, $I_{C+,-}(V_{PG})$ are shifted in positive/negative direction in $V_{PG}$, and $\Delta I_C(V_{PG})$ shows the same zero-field SDE whose polarity is switchable by $V_{PG}$.

Similar to Fig. 2e, we choose $V_{PG}$ such as $|\Delta I_C|$ is maximal and sweep $B$ in three orthogonal directions. The results are summarized in Supplementary Fig. 5 for both D1 and D2. In all the directions and on both D1 and D2, $|\Delta I_C|$ is maximal at zero field and decreases symmetrically at higher field. In all field directions and in two devices, $\Delta I_C(B)$ is an even function with $B$ and does not change sign with $B$, similar to the zero-field SDE introduced in the main text without external magnetic field, ferromagnetism, or other spontaneous TRSB mechanism[14-20].

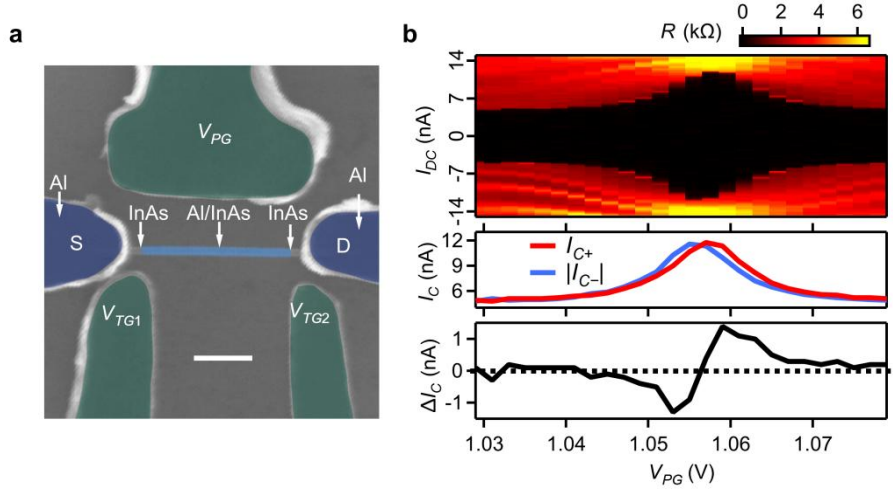

**Supplementary Fig. 4. Reproduced zero-field SDE in D2: a,** SEM image of D2 similar to D1. Scale bar: 500 nm. **b,** $R = dV/dI$ versus $V_{BG}$ and $I_{DC}$ (top panel). Extrated $I_{C+,-}(V_{PG})$ (middle panel) and $\Delta I_C(V_{PG})$ (bottom panel), showing similar SDE feature. $T \approx 10$ mK and $B = 0$. $V_{TG1}$ = 0 V, $V_{TG1}$ = 0 V, $V_{BG}$ = 6 V. $R_L$ = 11 kΩ is connected to D.

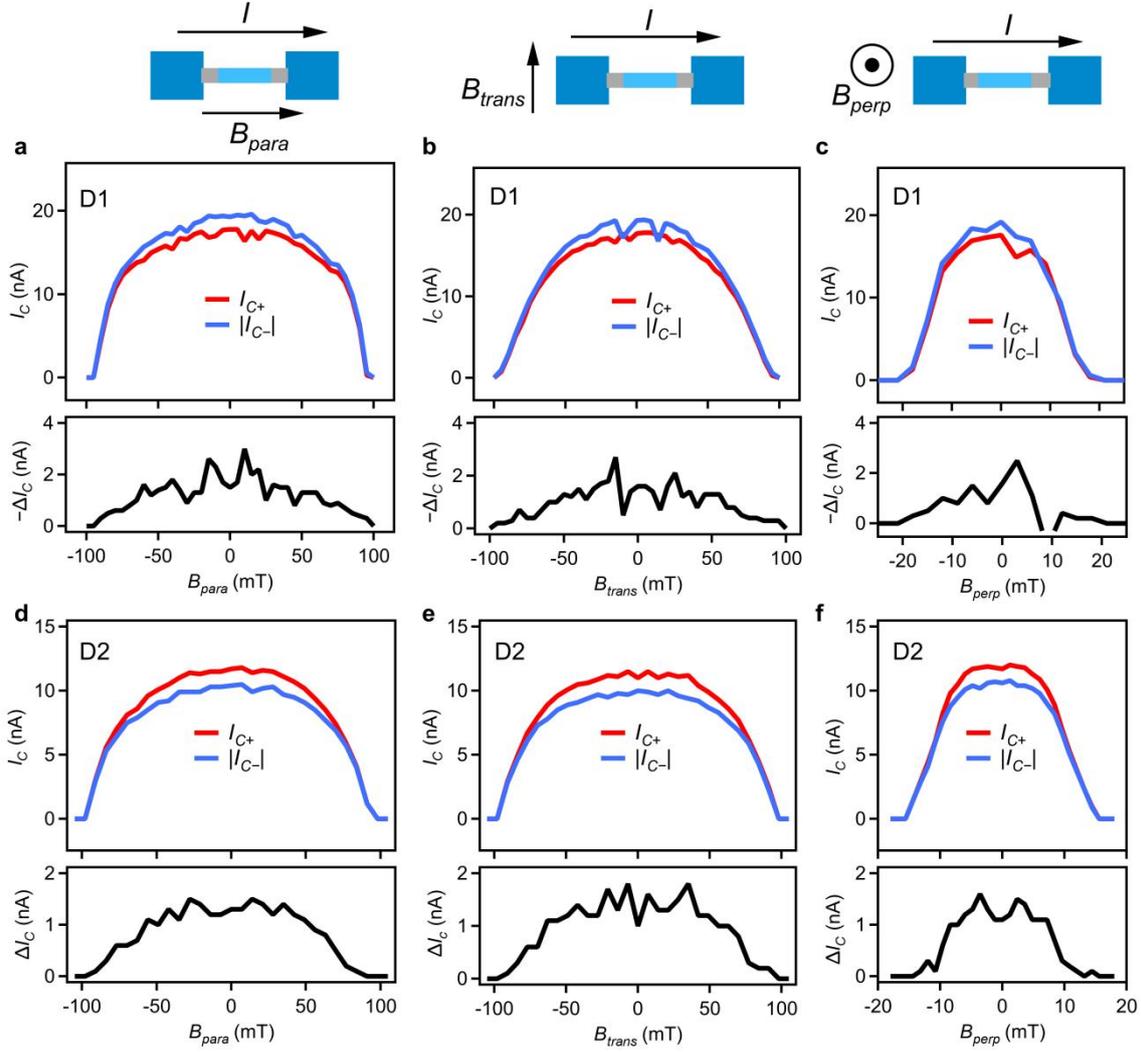

**Supplementary Fig. 5. Field dependence of SDE in three directions and in D1 and D2: a,** $I_{C+,-}$ (top panel) and $-\Delta I_C$ versus $B_{para}$ in parallel with the CPT (cartoon on top). **a** is reproduced from Fig. 2(E). $V_{PG}$ is such that $-\Delta I_C$ is maximal. **b,** Same as **a** but with $B_{trans}$ perpendicular to the nanowire and in parallel to the substrate (cartoon on top). **c,** Same as **a** but with $B_{perp}$ perpendicular to the substrate. **d-f,** Same as **a-c** but on D2. $V_{PG}$ is such that $\Delta I_C$ is maximal. $|\Delta I_C|$ is maximal at zero field and decreases symmetrically with finite field in all case.

## 5 Original dataset in Figs. 4g, h for D3

Supplementary Fig. 6 demonstrates how $dV_{PG}$ plotted in Figs. 4g, h is extracted from the original data. Taking the first plot in Supplementary Fig. 7b as an example (red and blue dots in Supplementary Fig. 6), $dV_{PG}$ is extracted by first interpolating the data (red and blue solid lines in Supplementary Fig. 6) and then calculating the gate difference between the maximal interpolated $I_{C+}$ and $|I_{C-}|$. All the other data of $dV_{PG}$ plotted in Figs. 4g, h are obtained from the original data in the same way.

Supplementary Figs. 7 plots the original dataset $I_{C+,-}(V_{PG})$, $\Delta I_C(V_{PG})$ and $\eta(V_{PG})$ for the extracted $dV_{PG}$ and $|\eta_{max}|$ in Figs. 4g,h with $R_L = 1$ k$\Omega$ (Supplementary Fig. 7a), 6 k$\Omega$ (Supplementary Figs. 7b,c), 11 k$\Omega$ (Supplementary Figs. 7d,e), respectively. Supplementary Figs. 8 plots the original dataset $I_{C+,-}(V_{PG})$, $\Delta I_C(V_{PG})$ and $\eta(V_{PG})$ for the extracted $dV_{PG}$ and $|\eta_{max}|$ in Figs. 4g, h, with $R_L = 16$ k$\Omega$ (Supplementary Figs. 8a,b), 21 k$\Omega$ (Supplementary Figs. 8c,d), 101 k$\Omega$ (Supplementary Figs. 8e), respectively.

Also, we note that all the original $I_{C+}(V_{PG})$ (and $I_{C-}(V_{PG})$) in Fig. 3d,e,g and all plots in Fig. 4 and Supplementary Figs. 7,8 have been subtracted by a small constant offset of -0.35 nA (and +0.35 nA) independent of $R_L$ and $I_C$ configurations. The original $I_{C+}(V_{PG})$ (and $I_{C-}(V_{PG})$) in Fig. 3i has been subtracted by an offset of -0.05 nA (and +0.05 nA). The original $I_{C+}(V_{PG})$ (and $I_{C-}(V_{PG})$) in Supplementary Fig. 3e has been subtracted by an offset of -0.23 nA (and +0.23 nA). Such small and constant offset is not due to the SDE. All the rest of the data in both the main text and Supplementary Information are without any subtraction.

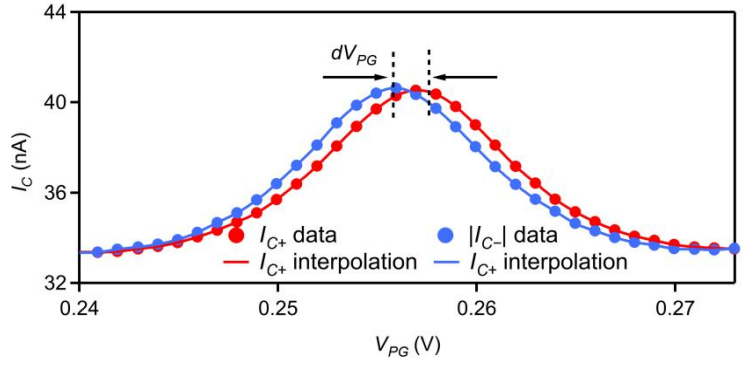

**Supplementary Fig. 6. Interpolation of data for accurate extraction of d$V_{PG}$:** Data $I_{C+}(V_{PG})$ (red dots) and $|I_{C-}(V_{PG})|$ (blue dots) are reproduced from the first plot in Supplementary Fig. 7b with interpolation (solid lines) enabling more accurate extraction of d$V_{PG}$ plotted in Figs. 4g, h.

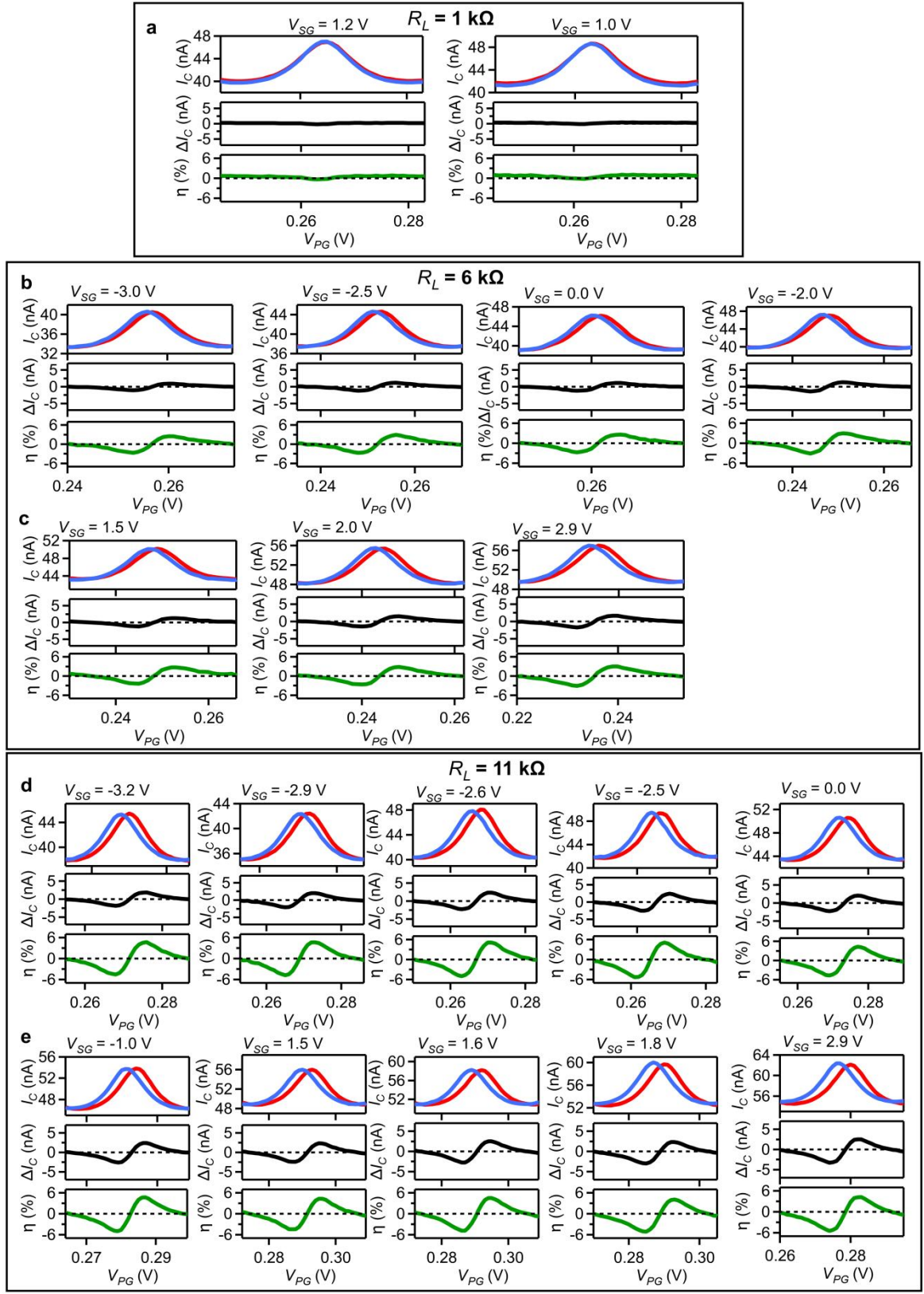

**Supplementary Fig. 7. Original dataset for Figs. 4g, h (1 kΩ, 6 kΩ, 11 kΩ): a-e,** $I_{C+,-}(V_{PG})$

(red, blue), $\Delta I_C(V_{PG})$ (black) and SDE efficiency $\eta(V_{PG}) = 2\Delta I_C/(I_{C+}+|I_{C-}|)$ (green) measured with D grounded and with different $R_L$ = 1 kΩ (**a**), 11 kΩ (**b,c**), 16 kΩ (**d,e**), respectively. Each figure is labeled with the corresponding $V_{SG}$. $V_{TG1}$ = -1.2 V, $V_{TG2}$ = -2.4 V, $V_{BG}$ = -3.5 V for all figures. The extracted d$V_{PG}$ and $|\eta_{max}|$ for all the configuration are plotted in Figs. 4g,h.

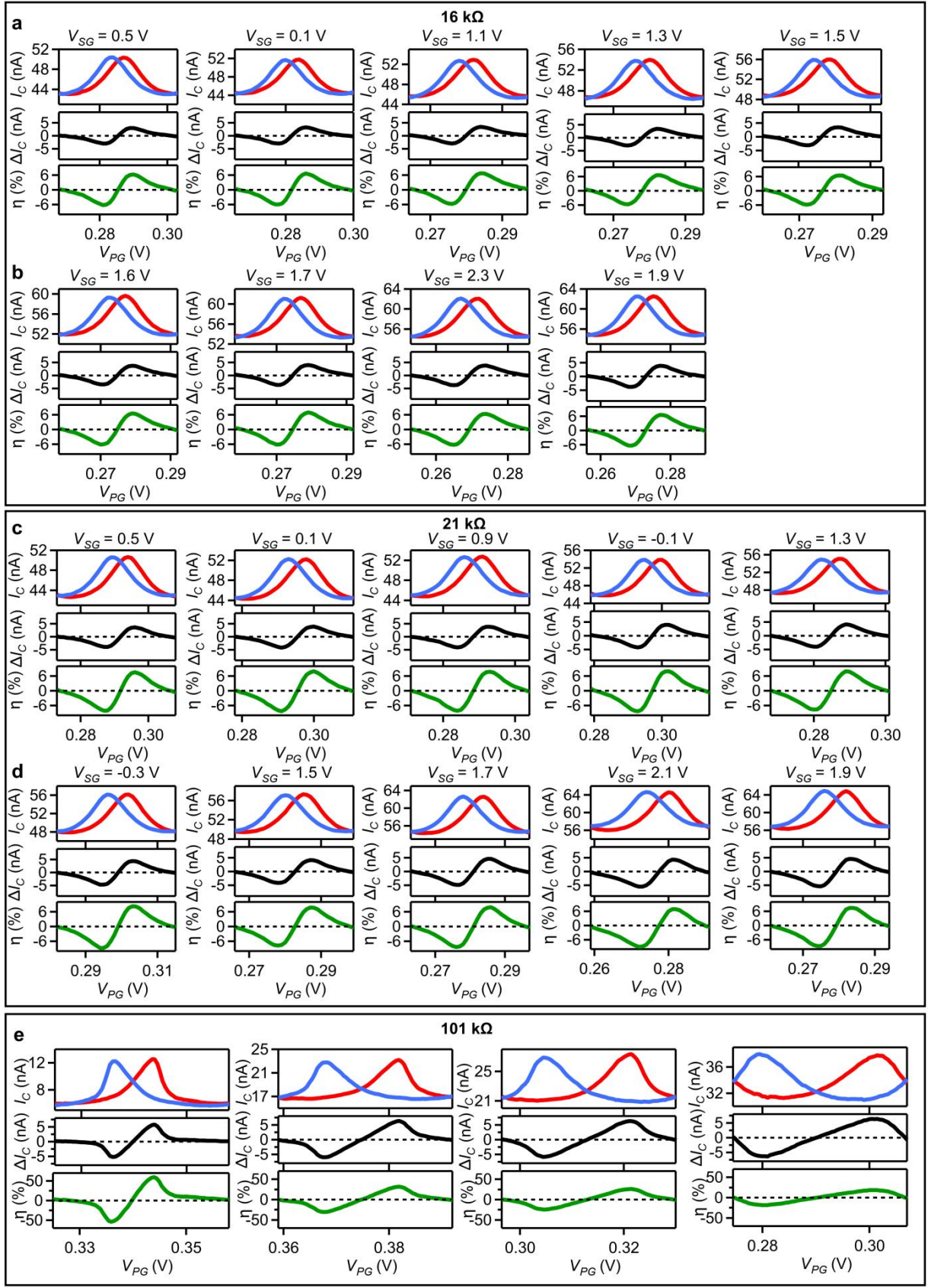

**Supplementary Fig. 8. Original dataset for Figs. 4g, h (16 kΩ, 21 kΩ, 101 kΩ): a-e, $I_{C+,-}$**

($V_{PG}$) (red, blue), $\Delta I_C(V_{PG})$ (black) and SDE efficiency $\eta(V_{PG}) = 2\Delta I_C/(I_{C+}+|I_{C-}|)$ (green) measured with D grounded and with different $R_L$ = 16 kΩ (**a,b**), 21 kΩ (**c,d**), 101 kΩ (**e**), respectively. **a-d:** Each figure is labeled with the corresponding $V_{SG}$. $V_{TG1}$ = -1.2 V, $V_{TG2}$ = -2.4 V, $V_{BG}$ = -3.5 V for all figures. **e:** from left to right: ($V_{BG}$ = -13.8 V, $V_{SG}$ = -3.0 V, $V_{TG1}$ = 4.5 V, $V_{TG2}$ = 1.4 V);  ($V_{BG}$ = -13.8 V, $V_{SG}$ = 0.0 V, $V_{TG1}$ = 4.5 V, $V_{TG2}$ = 1.4 V); ($V_{BG}$ = -13.8 V, $V_{SG}$ = 2.9 V, $V_{TG1}$ = 4.5 V, $V_{TG2}$ = 1.4 V); ($V_{BG}$ = -3.6 V, $V_{SG}$ = -3.3 V, $V_{TG1}$ = -1.2 V, $V_{TG2}$ = -2.5 V). The extracted d$V_{PG}$ and $\eta_{max}$ for all the configuration are plotted in Figs. 4g,h.

## 6 Theoretical model for $I_C(V_{PG})$ in CPT

With the CPT model in Fig. 1d, the quantum mechanical behaviors of the island are characterized by the conjugate charge and phase operators $\hat{n}$ and $\hat{\theta}$. Conventionally the Hamiltonian of CPT is constructed as[3,4,21]:

$$H = 4E_C(\hat{n} - n_g/2)^2 - E_J \cos(\hat{\theta} - \phi/2) - E_J \cos(\hat{\theta} + \phi/2) \quad \text{(S1)}$$

where $E_C = e^2/2C_\Sigma$ is the charging energy, and $E_J$ is the Josephson energy. The charge number $n_g = C_{PG}V_{PG}/|e|$ of the island and the phase difference across the transistor $\phi$ are treated as two experimentally controllable parameters. If the eigenenergies of (S1) are $E_n(n_g, \phi)$ ($n = 0, 1, 2,...$), the supercurrent is thus $I_S(n_g, \phi) = (2\pi/\Phi_0)\sum_n f(E_n)\partial E_n/\partial\phi$, where $f(E)$ is the Fermi-Dirac distribution. At base temperature, only the ground state has $f(E_0) \approx 1$ and all other $f(E_n) \approx 0$, thus $I_S(n_g, \phi) \approx (2\pi/\Phi_0)\partial E_0/\partial\phi$. The critical current $I_C(n_g) = \max_{0 \leq \phi \leq 2\pi}\{I_S(n_g, \phi)\}$, and the best fit to the experimental data can be used to extract the system's parameters $E_J$, $E_C$. The chemical potential shift is not included in the above Hamiltonian and the TRSB and ISB is also not present.

The diagonalization of (S1) can be calculated in the $\hat{n}$ representation[3,4,21], with the infinite number of basis $|N\rangle$ ($N$ is integer), representing a series of Cooper pair states. For numerical calculation, we take the maximal $N_{max} = 20$ which produces sufficient accuracy. A typical $E_{0,1}(n_g, \phi=0)$ is shown in Supplementary Fig. 9a (solid curves) in the regime of $E_C \sim E_J$ relevant to the main text. The oscillation with $|2e|$ period reflects the Coulomb blockade effect of Cooper pairs[3,4,21,22]. $E_0$ can thus be approximated by a series of states with fixed Cooper pairs number and $E_C$ when $n_g$ is around even integers (dashed curves in Supplementary Fig. 9a), while the neighboring states are hybridized due to $E_J$ when $n_g$ is around odd integers. Supplementary Fig. 9b show the fitting result (solid lines) to the typical $I_{C+}(n_g)$ data (Fig. 2c), with $E_C \approx 41$ μeV and $E_J \approx 80$ μeV.

In Figs. 4d-f with very large $I_C R_L$, we observe a significant tilt of $I_C(V_{PG})$, which cannot be reproduced from in the above simple Hamiltonian. We note that $I_C(n_g)$ is always symmetric with each Coulomb resonance and is not tilted even with significant barrier asymmetry[21,22]. However, we are able to explain such tilt also by $I_C R_L$. Let $I_C = g(n_g)$ where $g$ is their functional dependence, the shift $dn_g$ (or equivalently $dV_{PG}$ in Figs. 4d-f) is thus $dn_g = kg(n_g)$. Inserting this back iterately to $g$ thus gives the actual $I_C = g(n_g\text{-}dn_g) = g(n_g\text{-}kg(n_g\text{-}kg(n_g\text{-}...)))$. With small $R_L$ and small $I_C$ oscillation on the large background, $kg(n_g)$ can be approximated as the $n_g$-independent constant, which shifts $I_C = g(n_g)$ horizontally as in Fig. 1f and Supplementary Fig. 9c. However, with large $R_L$ and strongly oscillating $I_C$, the shift $kg(n_g)$ is also highly $n_g$-dependent. To illustrate this, we take a simplified functional dependence $I_C = g(n_g) = I_{C0}$ - $dI_C|\cos(\pi n_g/2)|$ as an example and iterate for 3 times (see caption). The result is plotted in Supplementary Fig. 9d showing qualitatively similar tilt in Figs. 4d-f.

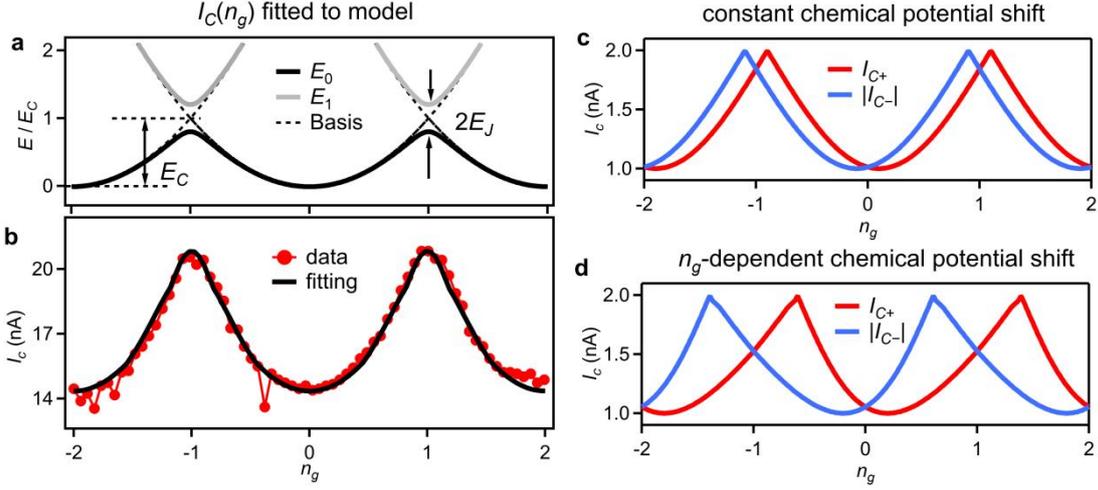

**Supplementary Fig. 9. Theoretical model for $I_C(V_{PG})$ in CPT: a,** The typical ground state energies $E_{0,1}(n_g, \phi=0)$ diagonalized by (S1) (solid curves). The energies for each unhybridized basis are drawn as the dashed curves (see text). **b,** Data $I_{C+}(n_g)$ (red dots) reproduced from Fig. 2(C) with the fitting (black line). $E_C \approx 41$ µeV and $E_J \approx 80$ µeV. **c,** Calculated $I_{C+} = g(n_g - 2dn_g/\pi) = I_{C0} - dI_C|\cos(\pi n_g/2 - dn_g)|$ (red) and $I_{C-} = -g(n_g + 2dn_g/\pi) = -I_{C0} + dI_C|\cos(\pi n_g/2 + dn_g)|$ (blue) with constant $dn_g = 0.1$. $I_{C0} = 2$, $dI_C = 1$. **d,** Calculated $I_{C+} = g(n_g - kg(n_g - kg(n_g - kg(n_g))))$ (red) and $I_{C-} = -g(n_g + kg(n_g + kg(n_g + kg(n_g))))$ (blue) with $k = 0.2$. $g(n_g) = I_{C0} - dI_C|\cos(n_g)|$ with $I_{C0} = 2$, $dI_C = 1$.

## 7 Coulomb diamond measurement of D1-D3

The Coulomb blockade effects in CPT devices described in Supplementary Information Sec. 6 is also commonly characterized by the "Coulomb diamond"[22-24] measurement by varying the dc bias voltage $V_{DC}$ and $V_{PG}$ and measuring the differential conductance $G = \mathrm{d}I/\mathrm{d}V$. We note that this is different from the SDE measurement where the dc current bias $I_{DC}$ is used and the differential resistance $R$ is measured. The results for D1-3 are summarized in Supplementary Figs. 10a-c, respectively. For D1, at $V_{DC} = 0$ inside the superconducting gap (bottom panel of Supplementary Fig. 10a), $G(V_{PG})$ oscillates with |2e| periodicity, is high on Coulomb resonance, and is low at Coulomb blockade[22-24]. When $V_{DC}$ is larger than the gap (middle panel of Supplementary Fig. 10a), the oscillation periodic of $G(V_{PG})$ is halved, reflecting the |1e| transport due to quasiparticles[22-24]. By comparison the oscillation periodicity between high and zero bias voltages, we also confirm that the CPT here operates in the even-parity regime[24]. The complete plot of $G(V_{DC,}\ V_{PG})$ shows a regular diamond shape (marked by the red dashed lines in Supplementary Fig. 10a, top panel), commonly called as the "Coulomb diamond"[22-24]. As a standard procedure[22-24], the charging energy $E_C = \mathrm{e}^2/2C_\Sigma$ can be estimated by such diamond, whose total height is $16E_C$ in the |2e|-oscillation regime (noted in Supplementary Figs. 10a-c, top panel). For D1, $E_C \approx 34$ μeV similar to that estimated by $I_C(V_{PG})$ in Supplementary Information Sec. 6. For D2 and D3, similar Coulomb oscillations confirm the even-parity regime[24] of D2 and D3. The similar diamond behaviors are also observed (Supplementary Figs. 10b, c) with the extracted $E_C \approx 11$ μeV (D2) and $E_C \approx 21$ μeV (D3). The in-plane field of $B = 0.2$ T is applied in D1 and D3 to suppress the superconductivity in the Al contacts while the island is still superconducting, so that the diamond structure is clearer. The estimated $E_C$ may be smaller than the one at zero field[22-24].

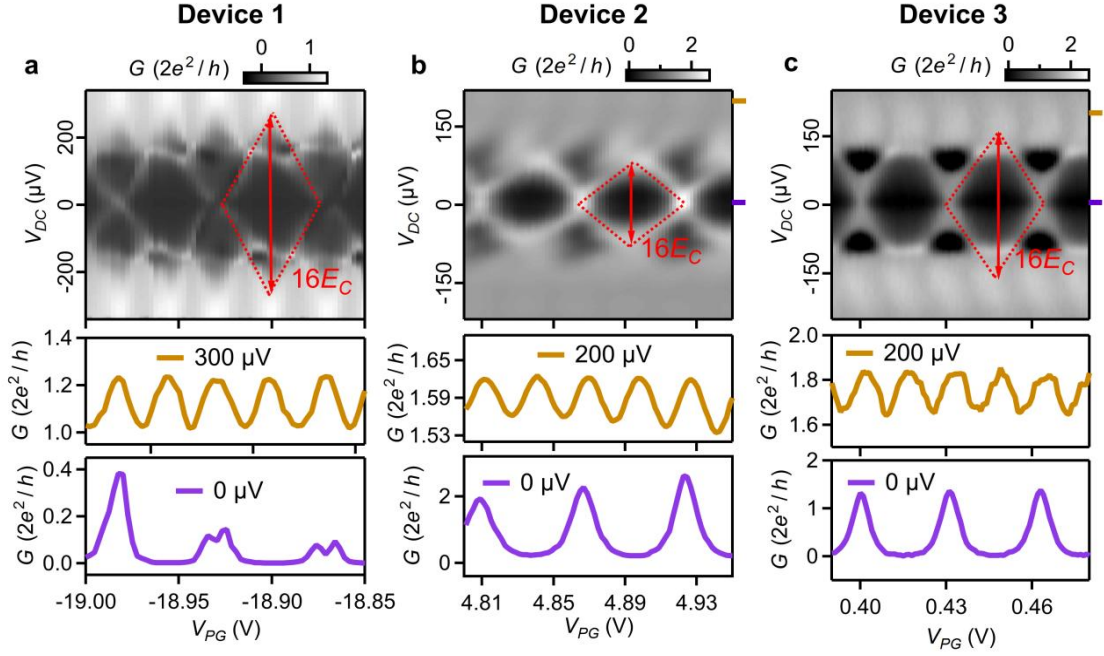

**Supplementary Fig. 10. Coulomb diamond measurement of D1-D3:** For D1: **a,** Top panel: $G = \mathrm{d}I/\mathrm{d}V$ versus $V_{PG}$ and $V_{DC}$, showing Coulomb diamonds (red dashed lines), with its height as $16E_C$ ($E_C$ is defined as $e^2/2C_\Sigma$) (*19*). $E_C \approx 34$ μeV. Middle and bottom panels: $G(V_{PG})$ for $V_{DC} = 0$ (showing |2e| periodicity) and $V_{DC} = 300$ μV (showing |1e| periodicity). $T \approx 10$ mK and in-plane $B = 0.2$ T. For D2: **b,** Similar to **a** with the estimated $E_C \approx 11$ μeV. $T \approx 10$ mK and $B = 0$. For D3: **c,** Similar to **a** with the estimated $E_C \approx 20$ μeV. $T \approx 10$ mK and in-plane $B = 0.2$T.